\documentclass[twocolumn]{aastex631}

\newcommand\aastex{AAS\TeX}

\newcommand{\swift}{\textit{Swift}}
\newcommand{\galex}{\textit{GALEX}}

\newcommand{\swim}{SwiM}

\usepackage{graphicx}	
\usepackage{amsmath}	
\usepackage{amssymb}	
\usepackage{longtable}
\usepackage{placeins}
\usepackage{mathtools}
\usepackage{float}
\usepackage[]{natbib}	
\usepackage[flushleft]{threeparttable}
\usepackage{enumitem}
\usepackage{hyperref}
\usepackage{verbatim}
\usepackage{multirow}
\accepted{August 15, 2023}

\submitjournal{ApJS}

\begin{document}

\title{\Large The New \textit{Swift}/UVOT+MaNGA (SwiM) Value-added Catalog}
\shorttitle{New SwiM Catalog}
\shortauthors{Molina et al.}

\correspondingauthor{Mallory Molina}
\email{mallory.e.molina@vanderbilt.edu}

\author[0000-0001-8440-3613]{Mallory Molina}
\affiliation{eXtreme Gravity Institute, Department of Physics, Montana State University, Bozeman, MT 59717, USA}
\affiliation{Department of Physics \& Astronomy, University of Utah, James Fletcher Building, 115 1400 E, Salt Lake City, UT 84112, USA}
\affiliation{Department of Physics \& Astronomy, Vanderbilt University, Nashville, TN 37240, USA}

\author[0000-0003-1752-679X]{Laura Duffy}
\affiliation{Department of Astronomy \& Astrophysics and Institute for Gravitation and the Cosmos, The Pennsylvania State University, University Park, PA 16802, USA}

\author[0000-0002-3719-940X]{Michael Eracleous}
\affiliation{Department of Astronomy \& Astrophysics and Institute for Gravitation and the Cosmos, The Pennsylvania State University, University Park, PA 16802, USA}

\author[0000-0001-9741-2703]{Mary Ogborn}
\affiliation{Department of Astronomy \& Astrophysics and Institute for Gravitation and the Cosmos, The Pennsylvania State University, University Park, PA 16802, USA}

\author[0000-0002-0863-1232]{Mary E. Kaldor}
\affiliation{Department of Astronomy \& Astrophysics, The Pennsylvania State University, University Park, PA 16802, USA}

\author[0000-0003-1025-1711]{Renbin Yan}
\affiliation{Department of Physics, The Chinese University of Hong Kong, Sha Tin, NT, Hong Kong, China}

\author[0000-0001-6842-2371]{Caryl Gronwall}
\affiliation{Department of Astronomy \& Astrophysics and Institute for Gravitation and the Cosmos, The Pennsylvania State University, University Park, PA 16802, USA}

\author[0000-0002-1328-0211]{Robin Ciardullo}
\affiliation{Department of Astronomy \& Astrophysics and Institute for Gravitation and the Cosmos, The Pennsylvania State University, University Park, PA 16802, USA}

\author[0000-0003-1469-8246]{Nikhil Ajgaonkar}
\affiliation{Department of Physics and Astronomy, University of Kentucky, 505 Rose St., Lexington, KY 40506-0057, USA}
\affiliation{Data Analytics with Layouts and Images, Intel Corporation, Hillsboro, OR}

\begin{abstract}

We present the the new \swift/UVOT+MaNGA (SwiM) catalog (SwiM\_v4.1). SwiM\_v4.1 is designed to study star-formation and dust attenuation within nearby galaxies given the unique overlap of \swift/UVOT near-ultraviolet (NUV) imaging and MaNGA integral field optical spectroscopy. SwiM\_v4.1 comprises 559 objects, $\sim4$ times more than the original SwiM catalog (SwiM\_v3.1), spans a redshift range $z\approx0.0002$--$0.1482$, and provides a more diverse and rich sample. Approximately 5\% of the final MaNGA sample is included in SwiM\_v4.1, and 42\% of the SwiM\_v4.1 galaxies are cross-listed with other well-known catalogs. We present the same data as SwiM\_v3.1, including UVOT images, SDSS images and MaNGA emission-line and spectral index maps with the same pixel size and angular resolution for each galaxy, and a file containing galaxy and observational properties. We designed SwiM\_v4.1 to be unbiased, which resulted in some objects having low signal-to-noise ratios in their MaNGA or \swift\ data. We addressed this by providing a new file containing the fraction of science-ready pixels in each MaNGA emission-line map, and the integrated flux and inverse variance for all three NUV filters. The uniform angular resolution and sampling in SwiM\_v4.1 will help answer a number of scientific questions, including constraining quenching and attenuation in the local Universe and studying the effects of black hole feedback. The galaxy maps, catalog files, and their associated data models are publicly released on the SDSS website\footnote{A description of the SwiM VAC is provided here: \url{https://www.sdss4.org/dr17/data_access/value-added-catalogs/?vac_id=swift-manga-value-added-catalog}, and the data are stored on the SDSS SAS: \url{https://data.sdss.org/sas/dr17/manga/swim/v4.1/}}.
\end{abstract}

\keywords{Catalogs (205) --- Galaxies (573) --- Astronomy data analysis (1858) --- Sloan photometry (1465) --- Ultraviolet photometry (1740) --- Spectroscopy (1558) --- Star formation (1569)}

\section{Introduction}\label{sec:intro}

Star formation growth and quenching are key elements of galaxy evolution, but their progression within galaxies is far from understood. In order to fully explore these processes, the appropriate star formation quenching models \citep[e.g.,][]{Martig2009,Springel2005,Steinhauser2016} and accurate dust attenuation laws \citep[e.g.,][]{Charlot2000,Battisti2016,Salim2018,swimdust,Zhou2022} must be used. The difficulty is that both of these functions depend on the spatial scale of the observations. Furthermore, different wavelength bands are sensitive to different star-formation timescales \citep[e.g., $\sim100$~Myr for the near-ultraviolet and $\sim10$~Myr for H$\alpha$; see][for details]{Kennicutt1998,Kennicutt2012,Calzetti2013}, and are affected by dust in different ways \citep[i.e.,][]{Calzetti2013}. Thus multi-wavelength data at the same angular resolution are necessary to accurately model both dust attenuation and construct star-formation histories within galaxies. In order to address these physical considerations, we constructed a sample of 150 galaxies containing both Sloan Digital Sky Survey IV (SDSS-IV) Mapping Nearby Galaxies at Apache Point Observatory  \citep[MaNGA;][]{Bundy2015,Yan2016,Blanton2017} optical integral field unit (IFU) spectroscopy and archival \textit{Swift} Ultraviolet Optical Telescope \citep[UVOT;][]{Roming2005} near-ultraviolet (NUV) images.  This dataset, called the {\it Swift}+MaNGA (SwiM) catalog, was first presented in \citet[][]{swimdr1}. The original SwiM catalog used version 3.1 of our data reduction pipeline, and will be referred to as SwiM\_v3.1 hereafter. The main data products in SwiM\_v3.1 were the ``maps'', which included \swift/UVOT NUV images, SDSS optical images, and MaNGA emission-line flux maps, equivalent width maps, and spectral index maps. All of the data were processed to have the same angular resolution and pixel size for each galaxy. Thus the SwiM catalog provided a unique and uniform view of the NUV and optical properties of nearby galaxies.

To date, our work using the SwiM\_v3.1 catalog has focused on dust attenuation. In \citet{swimdust} and Duffy et al. (submitted), we explored the relationship between the NUV slope, $\beta$, the nebular dust attenuation (or Balmer emission-line optical depth $\tau^l_B\equiv\tau^l_{{\rm H}\beta}-\tau^l_{{\rm H}\alpha}$), and infrared-excess (IRX\null). \citet{swimdust} found that the sight-line attenuation in the NUV can significantly contribute to the observed scatter in the $\beta$--$\tau^l_B$ relationship, while Duffy et al.\ (submitted) found that the IRX and $\beta$ measurements for star-forming regions are correlated with gas-phase metallicity. Recently, \cite{Zhou2022} also used the SwiM\_v3.1 catalog to study the spatial distribution of dust attenuation in nearby galaxies on kiloparsec scales. They found that both the slope and 2175~\AA\ ``bump'' in the attenuation curves are highly varied, but the strength of the 2175~\AA\ feature decreases with increasing specific star-formation rate (SFR), implying that local processes in star-forming regions heavily influence the shape of the attenuation curve. 

The analysis of the SwiM\_v3.1 catalog has thus already produced important insights into the properties of the attenuation laws that are relevant in the local universe. However, the small sample size of the original SwiM catalog limits the number of scientific questions that can be answered. For example, the lack of low-mass bluer galaxies and galaxies with active nuclei (AGNs) prevent detailed studies of AGN feedback and star-formation in the local Universe. Therefore, in order to expand on the work that can be done with the SwiM catalog, the number and types of galaxies included in the sample must be increased.

In this work, we present the new version of the SwiM catalog, which uses the final data release from the MaNGA collaboration and includes both archival and dedicated \swift/UVOT observations. The new SwiM catalog uses version 4.1 of our data reduction pipeline, and will be referred to as SwiM\_v4.1 hereafter. The new catalog comprises 559 objects, $\sim4$ times more than the first data release. SwiM\_v4.1 not only has better coverage of star-forming galaxies in both the high- and low-mass regimes, but also contains a significant number of galaxies with actively accreting black holes, providing a more rich and diverse set of galaxies than in the first data release. Thus, SwiM\_v4.1 will enable more in-depth studies of both individual objects and populations of galaxies, and is a powerful tool to study the physical processes that govern galactic evolution in the local Universe.

The sample selection of the SwiM\_v4.1 is presented in Section~\ref{sec:sample}. We describe the \textit{Swift}/UVOT and MaNGA data reduction in Section~\ref{sec:redux}, and the spatial matching of \swift\ and SDSS data in Section~\ref{sec:resamp}. The data products provided in SwiM\_v4.1 are described in Section~\ref{ssec:dr2prod}. We discuss the properties of the SwiM\_v4.1 catalog, its comparison to MaNGA and SwiM\_v3.1, and the volume-limited weight corrections in Section~\ref{sec:dr2prop}. The AGNs identified in in SwiM\_v4.1 is discussed in Section~\ref{sec:AGN}, and a summary is given in Section~\ref{sec:summ}.  We assume a $\Lambda$CDM cosmology when quoting physical properties such as masses, distances, and luminosities, with $\Omega_{\textrm{m}}=0.3$, $\Lambda = 0.7$ and $\textrm{H}_{0} = 70$~km~s\textsuperscript{$-1$}~Mpc\textsuperscript{$-1$}.

\section{{Defining the SwiM\_\MakeLowercase{v}4.1 catalog}}\label{sec:sample}
\subsection{SDSS-IV/MaNGA and Swift/UVOT}\label{ssec:survey}
The new SwiM catalog, SwiM\_v4.1, includes a subset of galaxies in the MaNGA survey \citep{Bundy2015,Yan2016}, which is an IFU spectroscopic survey included in the fourth generation of SDSS \citep[SDSS-IV;][]{Blanton2017}. The defining features of the MaNGA sample include: (1) a uniform stellar mass distribution for ${\rm M}_* > 10^9 {\rm M}_\odot$, as estimated via the SDSS $i$-band absolute magnitudes, (2) uniform spatial coverage in units of half-light radii ($R_e$), and (3) maximized spatial resolution and signal-to-noise ratio (S/N) for each galaxy \citep{Wake2017}. We used the MaNGA Product Launch 11 (MPL-11), which is the last data release, is presented in the SDSS-IV Data Release 17 \citep[i.e., SDSS--DR17;][]{sdssdr17}, and includes 10,145 galaxies.

While full descriptions of both the MaNGA and {\it Swift}/UVOT datasets are presented in \cite{swimdr1}, we summarize each of them here. The MANGA spectroscopic observations were obtained with hexagonal IFU fiber bundles mounted on the 2.5-m Sloan Foundation Telescope \citep{Gunn2006}, and fed into the dual-channel Baryon Oscillation Spectroscopic Survey (BOSS) spectrographs \citep{Smee2013}. The spectra span a total wavelength range from 3,622 to 10,350~\AA\ at a spectral resolution of $R\sim2000$. The data cubes have a point spread function (PSF) of $\sim2\farcs5$, and a spatial sampling of $0\farcs5$ with an average exposure time of $t_{\rm exp}\sim2.5$~hrs \citep{Yan2016}. However given that MaNGA is a ground-based survey, the measured PSFs for the individual datacubes are occasionally significantly larger than $2\farcs5$. We account for this in our sample selection as described in Section~\ref{ssec:sampsec}.

The \swift/UVOT is a 30-cm telescope, with a field of view (FOV) of $17^{\prime}\times17^{\prime}$ and an effective plate scale of $1^{\prime\prime}$ pixel$^{-1}$ \citep{Roming2005}.  The instrument has 3 NUV filters, two with wide-bandpasses (uvw2 and uvw1), and one with an intermediate-width bandpass (uvm2).  The three filters have slightly different PSFs, but all are close to $2\farcs5$, i.e., similar to that of the MaNGA optical spectra.  The properties of these filters are given in Table~\ref{table:sw_spec}.  As in the original catalog, none of our data suffer from coincidence loss \citep[see Section~3.3 in][for a thorough discussion]{swimdr1}.

\begin{deluxetable}{lccccc}
\tabletypesize{\footnotesize}
\tablecaption{\textit{Swift}/UVOT NUV Observation Properties}
\tablewidth{0pt}
\label{table:sw_spec}
\setlength{\tabcolsep}{4pt}
\tablehead{
{}& {Central}& {Spectral} & {PSF} & {Median} & {Minimum}\\
{ Filter} &{Wavelength} & {FWHM} & {FWHM} & {Exposure} & {Exposure}\\
{} & {(\AA)} & {(\AA)} & {(arcsec)} & {(s)} & {(s)}\\
{(1)} & (2) & (3) & (4) & (5) & (6)}
\startdata
{uvw2} & {1928} & {657} & {$2.92$} &{2460} & {135}\\
{uvm2} &  {2246} & {498} & {$2.45$} &{3043} & {104}\\
{uvw1} & {2600} & {693} & {$2.37$} &{1740} & {111}\\
\enddata
\tablecomments{Column 1: UVOT Filter, Columns 2--4: UVOT Filter properties as described in \cite{Breeveld2010}. The central wavelength for each filter assumes a flat spectrum in $f_{\nu}$. Columns 5--6: The median and minimum exposure times; the uvm2 statistics only include objects with uvm2 observations.} 
\end{deluxetable}

\subsection{SwiM\_v4.1 Sample Selection}\label{ssec:sampsec}
\begin{deluxetable}{lcccccc}
\tablecaption{Summary of New \textit{Swift}/UVOT NUV Observations}\label{table:sw_obs}
\setlength{\tabcolsep}{3pt}
\tablehead{
{MaNGA} & {R.A.} & {Dec.} & {Date Obs.} & {Date Obs.} & {$t_{\rm exp, w1}$} & {$t_{\rm exp, w2}$}\\
{ID} & {(deg)} & {(deg)} & {(uvw1)} & (uvw2) & (s) &(s)\\
(1) & (2) & (3) & (4) & (5) & (6) & (7)}
\startdata
1-23687 & 262.2574	& 60.0894 & 2020-07-31 & 2020-10-17 & 862 &892\\
1-38856 & ~54.8918 & $-$0.5119 & 2020-11-03 & 2020-11-08  & 1864 & 1839\\
1-71891 & 119.5533 & 38.5353 & 2020-12-07 & 2020-12-09 & 4346 & 3674\\
1-153077 & 119.2220 & 32.3456 & 2020-09-30 & 2020-09-21 & 2112 & 1922\\
1-153246 & 119.0444 & 33.2451 & 2020-12-06 & 2020-09-06 & 2011 & 3563\\
1-210866 & 245.8423 & 39.2153 & 2020-07-21 & 2020-07-24 & 890 & 861\\
1-351792 & 121.1725 & 50.7185 & 2020-11-26 & 2020-11-29 & \phantom{3}138\tablenotemark{a} & 823\\
1-352130 & 122.6799 & 52.5219 & 2020-08-24 & 2020-08-27 & 2663 & 1611\\
1-419427 & 182.2852 & 35.6358 & 2021-01-21 & 2021-01-22  & 1679 & 1930\\
1-457869 & 204.6236 & 26.0776 & 2020-11-10 & 2020-11-13  & 3530 & 2869\\
1-458124 & 203.5957 & 27.4613 & 2020-05-17 & 2021-02-10  & 2026 & 1612\\
1-592984 & 215.7184 & 40.6226 & 2020-12-01 & 2020-12-03  & 1951 & 1972\\
1-593748 & 229.3245 & 29.4005 & 2020-05-03 & 2020-05-10  & 819 & 903\\
\enddata
\tablecomments{
Column 1: MaNGA ID from MPL-11. Columns 2--3: Right ascension and declination, in degrees. Columns 4--5: The date of the first observation of each galaxy with the uvw1 and uvw2 filters, respectively. Columns 6--7: Total exposure time in seconds in the uvw1 and uvw2 filters, respectively.}\vspace{-2mm}
\tablenotetext{a}{The longer of the two uvw1 observations for 1-351792 did not have star tracking so we did not achieve the requested $\sim1$~ks exposure time.}
\end{deluxetable}
In order to define the SwiM\_v4.1 sample, we cross-matched the MPL-11 catalog with the UVOT data archive hosted by the High Energy Astrophysics Science Archive Research Center (HEASARC) as of August 2021, and required that each object was observed with both the uvw1 and uvw2 NUV filters. Of the 10,145 MPL-11 galaxies, 570 met these initial constraints. However, to further ensure both uniform and high-quality data across the entire SwiM catalog, we enforced a number of additional requirements for inclusion in the sample. Specifically, we included objects that had:
\begin{enumerate}[label=\arabic*),left=10pt,noitemsep]
    \item science-ready MaNGA data cubes in MPL-11,
    \item ${\rm PSF}_{\rm MaNGA}<{\rm PSF}_{\rm uvw2}$,
    \item both uvw1 and uvw2 exposures longer than 100~s 
    and images uncontaminated by foreground stars.
\end{enumerate}

While most of the data cubes in the MaNGA survey are science-ready, a small fraction suffered critical errors, or had many dead fibers, and thus are not usable. We removed 10 such objects from our sample. Additionally, 3 objects had MaNGA PSFs that were larger than that of the uvw2 filter. Since we base the final spatial resolution and sampling for all of the images and maps in the SwiM catalog on the resolution of the uvw2 data, we excluded these objects from consideration. Finally, we required that the exposure times for both UV filters be at least 100~s in duration, and that the images were not contaminated by bright foreground stars.  This excluded a further 11 objects from our sample. Thus, in total, we removed 24 galaxies from the dataset based on these four constraints, creating a final sample of 546 objects. We report the UVOT exposure time statistics in Table~\ref{table:sw_spec}. 

In addition to the wide-bandpass images, we also included the medium-bandpass uvm2 images in the final data cubes, when available. For our sample of 546 objects, 496 had archival uvm2 observations, 6 of which had exposure times of less than 100~s. We thus excluded the uvm2 images for those 6 objects, leaving a total of 490 objects with uvm2 observations. 

In addition to the 546 objects with archival data, we performed dedicated \swift/UVOT observations of a further 13 galaxies. The new observations were taken between May 2020 and January 2021 as part of the GI program 16190136 (P.I. Molina), with the goal of creating a catalog with similar properties to the MaNGA sample. Thus the final SwiM\_v4.1 catalog comprises 559 galaxies. We required that the new observations have sufficiently high S/N at large galactocentric radii ($\textrm{S/N} \gtrsim 10$ for star-forming galaxies and $\textrm{S/N} \gtrsim 5$ for systems currently being quenched), resulting in $\sim1\mbox{--}2$~ks per filter for each object. Since uvm2 was not required for inclusion in our sample, we only completed dedicated observations in the uvw2 and uvw1 filters. We summarize the new observations in Table~\ref{table:sw_obs}. 


\section{\swift\ and MaNGA Data Reduction}\label{sec:redux}
The final SwiM\_v4.1 catalog includes SDSS and {\it Swift}/UVOT imaging, as well as MaNGA emission line, equivalent width, and Lick indices maps, all of which have been transformed to have the same spatial resolution and sampling of the galaxy's uvw2 image. This requires a multi-step reduction process which is completely described in Sections~3 and 5 of \cite{swimdr1}. Here we provide a brief overview of the data reduction for the \swift\ images and MaNGA data cubes, which includes a detailed description of the updated \swift\ UVOT Pipeline. We describe the spatial resolution matching and re-sampling processes in Section~\ref{sec:resamp}.

\subsection{\textit{Swift}/UVOT Pipeline}
\label{ssec:sw_pipeline}
We processed the raw UVOT data from HEASARC using the \swift\ UVOT Pipeline\footnote{\url{github.com/malmolina/Swift-UVOT-Pipeline}}, which is an improved and automated version of the subroutine \texttt{uvot\_deep.py} written by Lea Hagen\footnote{\url{github.com/lea-hagen/uvot-mosaic}}. The pipeline was originally described in detail in \cite{swimdr1}, and has been updated for this work. The new pipeline automates a modified version of the \texttt{uvot\_deep.py} subroutine and creates multiple science-ready mosaics in a single execution. One of the notable new features introduced in the updated version of the pipeline is a re-binning subroutine, which allows for a more complete use of the UVOT archive. We briefly describe the data reduction process and highlight the updated changes to the pipeline below.

The UVOT data are reduced following the data processing procedures outlined in the UVOT Software Guide\footnote{\url{heasarc.gsfc.nasa.gov/docs/swift/analysis}}. Each UVOT image is a summation of a series of $\sim11$~ms frames taken over the specified observation time. These images are then stacked to create the final image mosaic. 

The individual images and exposure maps are first aspect-corrected, i.e., aligning the individual $\sim11$~ms frames to the position of known sources. The counts images are then corrected for large-scale sensitivity issues associated with the detector, and all bad pixels are masked. Occasionally, \texttt{uvot\_deep.py} will produce errors that cause the exposure map to have 0 or NaN values for small regions. This error only occured for one galaxy in our sample and does not affect the galaxy itself. We provide masks to correct this issue, as described in Appendix~\ref{app:swimmap}.

The updated version of the pipeline then corrects each image for the degradation of the detector as described in the 2020 update of the UVOT Software Guide\footnote{\url{https://heasarc.gsfc.nasa.gov/docs/heasarc/caldb/swift/docs/uvot/uvotcaldb_throughput_06.pdf}}, as well as the dead time which accounts for approximately 2\% of the full frame's exposure time. 

The UVOT software then requires that each image included in the stacked mosaic be aspect-corrected, have a single frame time (currently defined as the standard full frame exposure time of 11.0322~ms), and be $2\times2$ binned to produce a plate scale of $1^{\prime\prime}$ pixel$^{-1}$. The last step involves binning groups of 4 pixels and is usually completed by the UVOT on-board processing before the data are sent down. However, there are instances where images in the archive are $1\times1$ binned. The new version of the pipeline re-bins these images and updates the world coordinate system (WCS) accordingly, thus allowing the frames to be used in the mosaic. After completing these data quality checks, all science-ready images are stacked to create the final mosaic. 

We note that cosmic ray corrections are not necessary for UVOT images given that the images are the sums of $\sim11$~ms frames. In this observing regime, any cosmic ray that hits the detector will only affect a few frames and will produce only a few counts in a single location. The hit will therefore be incorporated into the sky background counts of the summed image. 
 
Finally, coincidence loss, which occurs when two or more photons arrive at a similar location within the same $\sim11$~ms frame, is negligible for all the galaxies in our sample.  In particular, for extended sources, coincidence loss can be neglected when the count rates is less than 10 counts~s$^{-1}$~pixel$^{-1}$ \citep{Breeveld2010}. For comparison, the maximum count rate across all the UVOT observations in our sample is $\sim2.1$~counts~s$^{-1}$~pixel$^{-1}$, which implies a correction factor of $<0.3$\%.  This is significantly smaller than the dead-time correction of 2\%, allowing us to ignore the effects of coincidence loss.  A more thorough description of coincidence loss is provided in Section~3.3 in \cite{swimdr1}. 

\subsection{\textit{Swift}/UVOT Sky Subtraction}
\begin{deluxetable}{ll}
\tablecaption{Galaxies with \textit{Swift} Exposure Time Differences}
\label{table:sw_edge}
\setlength{\tabcolsep}{35pt}
\tablehead{
{MaNGA ID}&{Issue Location}\\
{\phantom{10}(1)} & \hspace{6mm}(2)}
\startdata
1-33510 & Background\\
1-76706 & Background\\
1-117998 & Background\\
1-301193 & Background\\
1-321979 & Galaxy\\
1-324762 & Galaxy\\
\enddata
\tablecomments{Column 1: MaNGA ID, Column 2: Location in the \textit{Swift}/UVOT images where exposure time differs from the center of the galaxy of interest.}
\end{deluxetable}
While a detailed description of the UVOT sky subtraction is given in Section~3.4 of \cite{swimdr1}, we briefly summarize the procedure here. We calculate the sky background counts using an elliptical annulus, where the inner radius is twice the Petrosian semi-major axis \citep[$R_p$ in the NASA-Sloan Atlas;][]{Blanton2011}, and the outer circular aperture radius is $4R_p$. We then mask all astrophysical contaminants in the sky annulus using Source Extractor \citep{Bertin1996}, as well as any pixels that do not have the same exposure time as the center of the galaxy of interest. This issue affects the six galaxies in our sample listed in Table~\ref{table:sw_edge}: four where significant regions of the surrounding background have different exposure times (denoted as ``background'' in Table~\ref{table:sw_edge}), and two where different regions of the galaxies themselves have different exposure times (denoted as ``galaxy'' in Table~\ref{table:sw_edge}). We urge caution when using the data for these galaxies and provide the original non-sky-subtracted \swift\ data in the catalog for all galaxies in the sample.  Finally, we use the biweight estimator from \cite{Beers1990} to calculate the final background counts for the object. The background counts for each galaxy in each filter are given by the keywords \texttt{SKY\_W1}, \texttt{SKY\_M2} and \texttt{SKY\_W2} in the header data unit (HDU) 17 for each data cube.

\subsection{MaNGA Data Reduction and Processing}
We use the fully reduced MaNGA emission-line flux, equivalent width (EW), and spectral indices maps that are presented in MPL-11 and are identical to that of SDSS-DR17 \citep{sdssdr17}. These finalized data products are created via a two-step process. First, all MaNGA spectra are reduced using the updated MaNGA Data Reduction Pipeline \citep[DRP;][]{Law2021}, which processes the raw data and creates flux-calibrated, sky-subtracted, co-added data cubes for each galaxy. After this initial reduction, the spectra are then fed into the MaNGA Data Analysis Pipeline \citep[DAP;][]{Westfall2019}. The DAP takes the reduced IFU spectral data cubes and constructs 2--D maps, including the flux, equivalent width, and indices maps that we incorporate into our catalog. The DAP also provides measured quantities such as emission-line fluxes for various apertures within the galaxy.

The MaNGA spectra are fit by the DAP using multiple analysis methods which are differentiated by the keyword ``DAPTYPE''. We adopt the ``HYB10-MILESHC-MASTARHC2'' analysis method for two reasons: 1) the ``HYB10'' scheme is optimized for emission-line measurements, and 2) the results from this analysis method were available for all of the galaxies in the SwiM\_v4.1 catalog. 

A complete description of the spectral fitting process is given in Section~5 of \cite{Westfall2019} and Section~4.2 of \cite{Law2021}; we only briefly describe the process here. First the individual spaxels are Voronoi-binned \citep{Cappellari2003}, and the stellar continua are fit with the penalized pixel-fitting code \citep[pPXF;][]{Cappellari2004} that uses a combination of stellar templates from the MILES stellar library \citep{Sanchez2006}. This initial fit fixes the stellar kinematics for subsequent fits. Next, the entire spectrum for each individual spaxel (both emission lines and stellar continuum) is fit with pPXF using a combination of stellar templates presented in \cite{Maraston2020}, which are derived from the MaNGA Stellar Library \citep[MaStar\footnote{\url{https://www.icg.port.ac.uk/MaStar/}};][]{Yan2019}. The DAP uses the MaStar template library, rather than the MILES library, since the former covers the entire MaNGA spectral range.  This allows all the data to be analyzed and enables more lines to be fit, as described in Section~4.2.1 of \cite{Law2021}. We note that we recalculate both the Lick indices and $D_n(4000)$ maps to allow for binned measurements, as described in Section~\ref{sec:resamp}.

\subsection{Extinction Corrections in SwiM\_v4.1}
The SDSS and \swift\ images are {\it not} corrected for either foreground extinction or internal attenuation. However, the MaNGA maps are corrected for foreground extinction by the DAP, which uses the \cite{Schlegel1998} $E(B-V)$ value and assumes the Milky Way extinction curve from \cite{Odonnell1994}, with $R_V=3.1$. We keep this foreground extinction correction in our maps. Finally, the SFR measurements provided in the Swim\_all catalog file are additionally corrected for internal attenuation, assuming the \cite{Odonnell1994} law, ${\rm R}_{V}=3.1$ and an H$\alpha$/H$\beta$ ratio of 2.86 \citep[][chapter 11]{osterbrock2006}. See Section~\ref{ssec:swim_cat} for more details on the SFR calculation.

\section{Spatially Matching SDSS and \swift\ Data}\label{sec:resamp}

Since the goal of the SwiM\_v4.1 catalog is to facilitate a joint analysis of SDSS and \swift\ data, we transform all images and maps to the same spatial resolution and sampling. We adopt the sampling and resolution of the uvw2 image, as it has the coarsest PSF (2\farcs92). A detailed description of the transformation process for each dataset in the catalog is given in Section~5 of \cite{swimdr1}. However, we provide a brief overview of the processes here.

\subsection{\swift/UVOT and SDSS Imaging}\label{ssec:image_resamp}
In order to transform the uvm2, uvw1, $u$, $g$, $r$, $i$, and $z$ images of \swift\ and SDSS to the resolution of uvw2, we first convolve each image with a 2--D Gaussian kernel whose standard deviation is given in equation (1) in \citet{swimdr1}. We assume the PSFs for the \swift\ filters have the full-width at half maximum (FWHM) presented in column 4 of Table~\ref{table:sw_spec}, while the SDSS images have a PSF FWHM of 1\farcs4.

After the images are degraded to the same resolution as uvw2, we reproject the data onto the same WCS and spatial sampling using the \texttt{reproject.reproject\_exact} function in \texttt{Astropy} \citep{Whelan2018}. We compensate for any broadening in the \swift\ images associated with the reprojection process by applying a correction factor, $\epsilon$, which is a polynomial that depends on the fractional pixel shift between the pixel grids. The $\epsilon$ correction accurately estimates the broadening effect to less than 0\farcs001, which is significantly smaller than the 2\farcs92 PSF of uvw2. The broadening effect is much smaller for SDSS data, so we adopt a median correction for all of the SDSS images; this results in an error that is $<1\%$ of the final PSF width. 

The entire process described above is also applied to the exposure maps and masks. All masked pixels were ignored in the computation. For the new (and in the case of SDSS, larger) pixels, if more than 40\% of the new pixel area comes from bad pixels then the final pixel is masked.

We also account for the covariance introduced by the convolution and reprojection processes. In lieu of providing covariance matrices, we give the functional forms for $f_{\rm covar}=\sigma_{\rm covar}/\sigma_{\rm no\_covar}$ in equations (3)--(5) in Section~5.2 of \citet{swimdr1}. 

\subsection{MaNGA Emission-Line Flux and Equivalent Width Maps}\label{ssec:flux_samp}
Since the EW maps are the ratio of the line flux and continuum maps, they must be deconstructed to accurately transform them to the same spatial resolution and sampling as uvw2. We create the continuum maps by simply taking the ratio of the emission-line flux and EW maps. After this step, we apply the same convolution and reprojection process described in Section~\ref{ssec:image_resamp} to both the MaNGA emission-line flux and continuum maps. All masked pixels are again ignored in the computation, and each pixel in the mask map is rounded to 0 or 1 to create the final mask. When calculating binned EWs from the final maps, we recommend that the user bin the flux and continuum maps separately before taking the ratio to compute the EW.

Unlike the small broadening effect in the direct images, there is significant covariance in the MaNGA maps. To account for this variation, we multiplied the pixel correlation matrix by the refitted variance maps to build a final covariance matrix. We then distill this information into a functional form for $f_{\rm covar}$, which is given by equation (11) in \citet{swimdr1}. 

In addition to the errors cited above, we note that testing by \citet{Belfiore2019} demonstrated that the H$\alpha$ and EW errors are larger than the formal error by 25\%. Therefore, users should multiply their errors by 1.25 for more realistic uncertainty estimates.

\subsection{MaNGA Spectral Index Maps}
\subsubsection{D$_n$(4000)}
Given the different definition and units of D$_n$(4000) compared to the rest of the spectral indices, we have a separate routine for its calculation and present it in a different HDU in the final catalog. We begin by re-measuring the blue and red band flux densities in the DRP \texttt{LOGCUBE} files, and then apply the same convolution, reprojection and covariance processes described in Section~\ref{ssec:flux_samp}. We process the variance and mask maps the same way, and present all of these data in a separate HDU in the final data cube. 

When calculating binned D$_n$(4000) measurements, the user should bin the red and blue flux bins independently and follow the directions in Appendix~\ref{app:swimmap} and Equation~\ref{eqn:d4000_sig} to calculate both D$_n$(4000) and its associated uncertainty. The same covariance scale factor, equation (11) in \citet{swimdr1}, should be applied to the errors, in addition to a scale factor of 1.4 \citep{Westfall2019}. 

\subsubsection{Lick Indices}
The rest of the spectral index maps are created using the same procedure. We begin by subtracting the best-fit emission-line spectrum from each spaxel in the MaNGA DRP \texttt{LOGCUBE} data cube and then transform it to the rest frame with the redshift and stellar velocity given by the MaNGA DAP\null.  For each Lick index, we then measure each spectral index band's integrated flux and continuum flux density, using the passbands from the MaNGA DAP and equations (6)--(10) in Section~5.4 of \citet{swimdr1}. We then use the same convolution, reprojection and covariance processes presented in Section~\ref{ssec:flux_samp} to the flux, continuum, uncertainty and mask maps. 

In order to get binned Lick index values, the user should bin the continuum and flux maps separately before using equation~\ref{eqn:app_spidx} given in Appendix~\ref{app:swimmap}. The spectral window, $\Delta\lambda$, for each spectral index is provided in Table~\ref{table:specindx_dm}. Finally, to estimate realistic errors, the user should apply the scale factor from equation (11) in \citet{swimdr1} to account for covariance and a scale factor of 1.2 for H$\beta$ and H$\delta_A$ absorption EW, 1.6 for the Fe5335 index, 1.4 for the Mgb index, and 1.5 for the NaD index. We note that we adopt the \citet{Burstein1984} Lick index definition in order to facilitate further binning of the final maps. 

Traditionally, the Lick-index system is defined with a constant instrumental resolution of ${\rm FWHM}=8.4$~\AA\ and at a fixed stellar velocity dispersion \citep[e.g.,][]{Worthey1997}. The flux, continuum and their uncertainty maps presented in the SwiM\_v4.1 catalog have {\it not} been transformed to have this uniform spectral resolution. Instead, we provide flux-weighted, combined ``dispersion'' maps for all the spectral indices, which include the effects of both instrumental resolution and stellar velocity dispersion. This approach allows the user the freedom to define their own uniform spectral resolution, by either manipulating the data or matching models to the spectral resolution of the data. We calculate the combined ``dispersion'' maps by adding the stellar velocity dispersion and instrumental spectral resolution in quadrature for each spaxel and each index, and also provide their associated uncertainty and mask maps.
\section{SwiM\_\MakeLowercase{v}4.1 Data Products}\label{ssec:dr2prod}
The SwiM\_v4.1 catalog consists of three main data products: 1) the SwiM\_all catalog file, 2) the maps and 3) the SwiM\_eline\_ratios, i.e., the emission-line detection file. The first two products are updated and modified versions of those provided in the first SwiM data release, while the third is new to this data release. We provide detailed descriptions of all three of these products in Appendices~\ref{app:swimcat}--\ref{app:swimeline}, and a brief description below. All of the data described below are publicly available on the SDSS website as a Value-Added catalog (VAC)\footnote{\url{https://data.sdss.org/sas/dr17/manga/swim/v4.1/}}.

\subsection{SwiM\_all catalog File} \label{ssec:swim_cat}
The description of the catalog file can be found in Appendix~\ref{app:swimcat}. This file holds all the basic physical properties, MaNGA and \swift\ observation properties, and integrated flux measurements for the galaxies in the sample. The only difference in format between the SwiM\_v3.1 and SwiM\_v4.1 catalog file is the addition of the DRP3QUAL value from the the MaNGA \texttt{drpall} file. DRP3QUAL is a bitmask that describes the quality of the data reduction. We also note that \texttt{NSA\_ELPETRO\_THETA} has been replaced with \texttt{NSA\_ELPETRO\_TH50} to maintain consistency with the latest MaNGA \texttt{drpall} file. More information on the DRP3QUAL can be found in the MaNGA DAP documentation\footnote{\url{https://sdss-mangadap.readthedocs.io/en/latest/metadatamodel.html}}. 

We note that the integrated \swift\ fluxes presented in the catalog file are aperture-corrected to the SDSS $r$-band; this maintains consistency with the integrated fluxes reported from the MaNGA \texttt{drpall} file. In the catalog file, we include the aperture correction factors in addition to the \swift\ integrated fluxes and inverse variances. We provide attenuation and foreground extinction-corrected star formation rate estimates within 1 effective radius, which are calculated by converting the foreground extinction-corrected H$\alpha$ emission measurement provided in the \texttt{dapall} file using the relation provided in \cite{Kennicutt2012}. We additionally correct for internal attenuation with the \cite{Odonnell1994} law, ${\rm R}_{V}=3.1$ and an assumed intrinsic H$\alpha$/H$\beta$ ratio of 2.86 \citep[see chapter 11 of][]{osterbrock2006}. For more details on the star formation rates and integrated flux calculations; see Sections~2.3 and 4 of \citet{swimdr1}, respectively.

Finally, we present the correction factors for the volume-limited weights for the galaxies in our sample. We note that not all of the galaxies in the SwiM\_v4.1 catalog are included in the MaNGA primary+full secondary samples \citep[see][for more detailed descriptions of the different MaNGA sample selections]{Wake2017}; these additional objects do not have correction factors. A more detailed discussion of the volume-limited weights is presented in Section~\ref{ssec:vlw_corr}.

\subsection{SwiM\_v4.1 Maps}
The maps are the main component of the SwiM catalog. Each galaxy has a set of maps containing all of the SDSS and \swift/UVOT data, which have been transformed to the same WCS, spatial resolution, and sampling as its uvw2 image. The only difference in format between the SwiM\_v3.1 and SwiM\_v4.1 maps is the addition of 13 new lines that were not measured in previous versions of the DAP\null. The data models for the maps can be found in Appendix~\ref{app:swimmap}.

\subsection{SwiM\_eline\_ratios File}
The final cataloged dataset is the SwiM\_eline\_ratios file, which is new to SwiM\_v4.1. In order to provide as much information as possible, we have included all emission-line maps for every galaxy, regardless of the amount of spaxels that are masked. As a result, there are some emission-line flux maps that have few science-ready spaxels in the final product. For each emission-line map of each galaxy, we store the ratio of unmasked-to-total spaxels within one elliptical Petrosian radius in the SwiM\_eline\_ratio file. The file provides a convenient way to determine which galaxies, and thus which maps, will be useful for the user to download. The data model for this file is provided in Appendix~\ref{app:swimeline}.


\section{Properties of the galaxy sample in the SwiM\_\MakeLowercase{v}4.1 catalog}\label{sec:dr2prop}

\begin{figure*}
\centering
\includegraphics[width=0.48\textwidth]{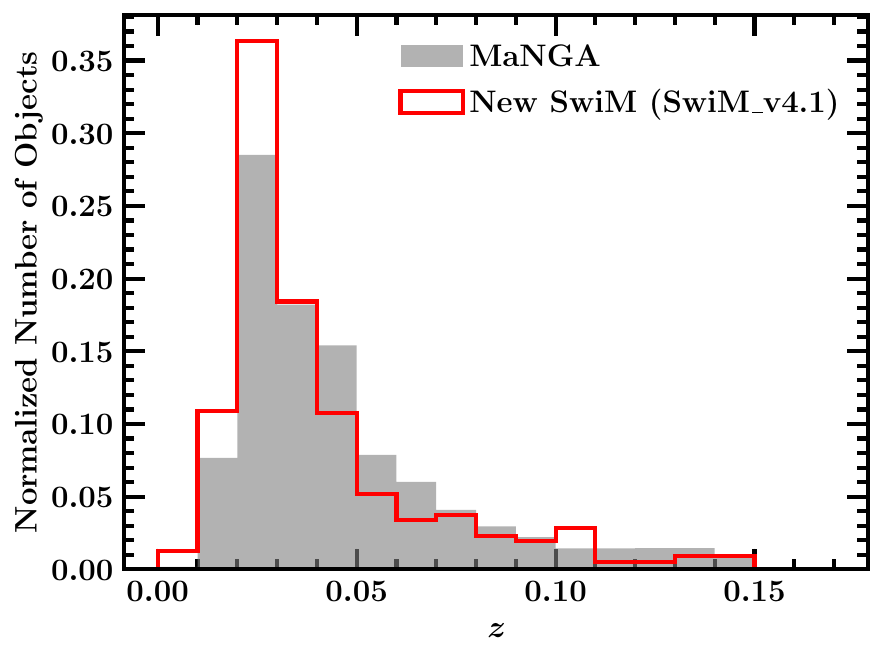}
\includegraphics[width=0.48\textwidth]{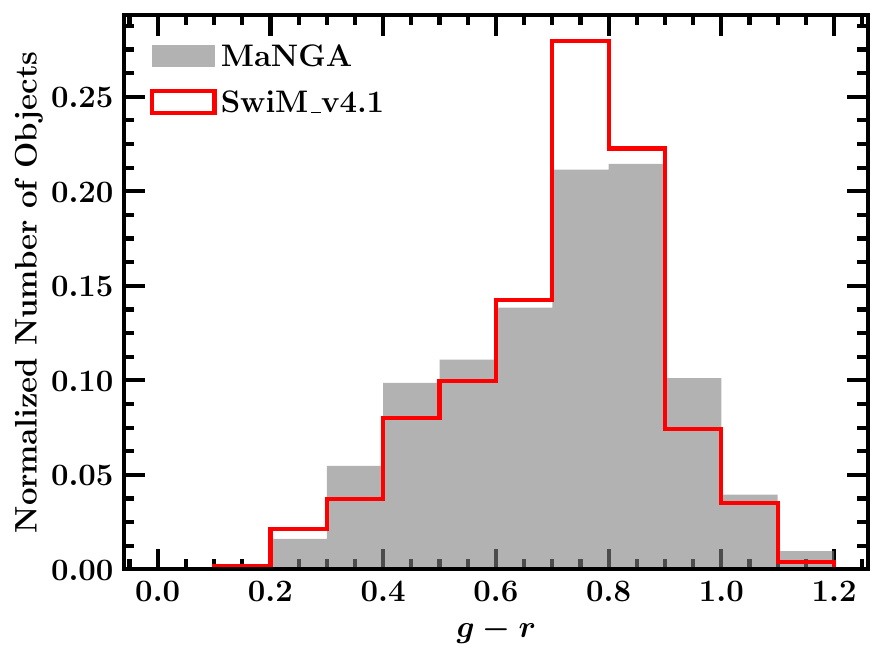}
\includegraphics[width=0.48\textwidth]{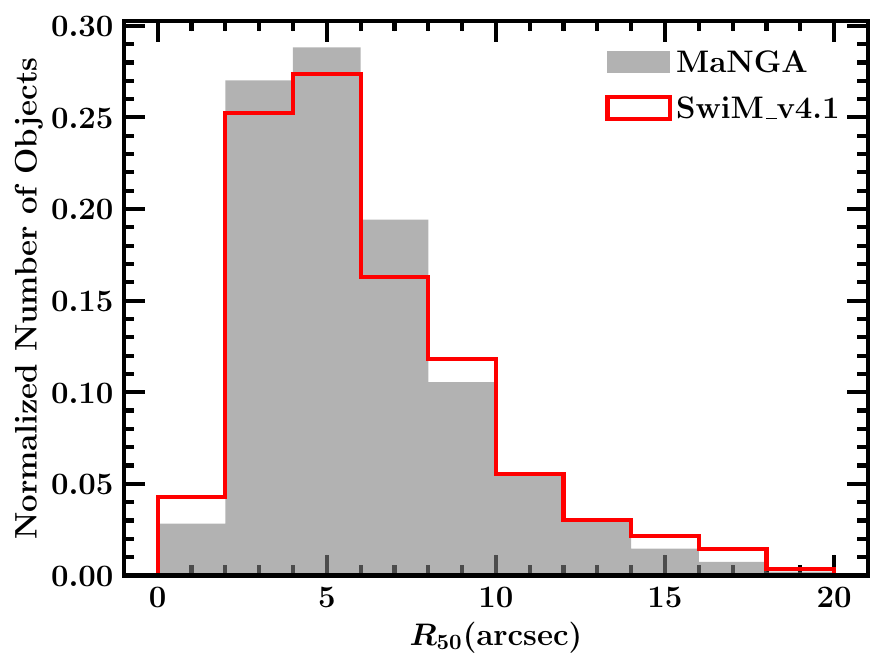}
\includegraphics[width=0.48\textwidth]{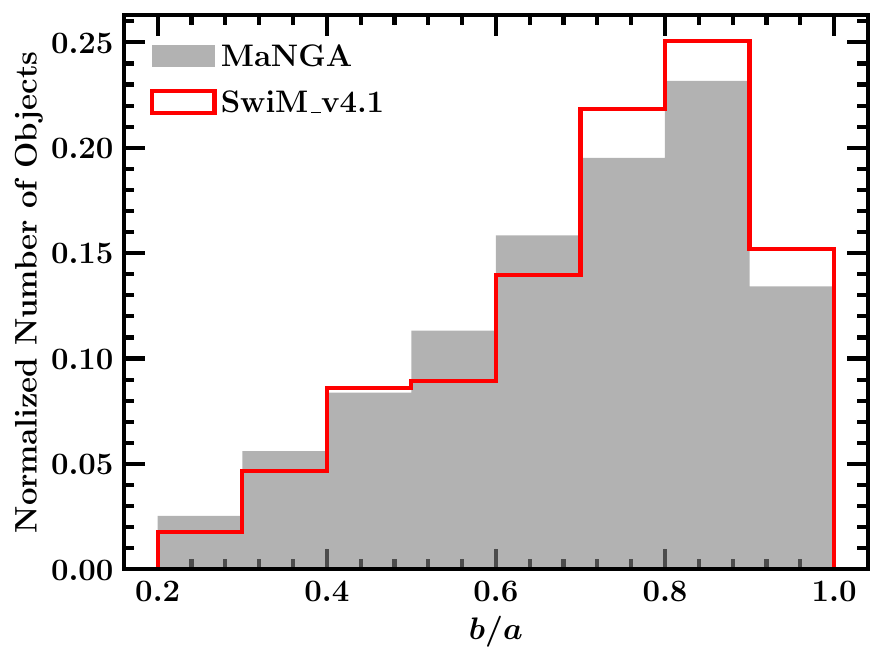}
\caption{Distributions of redshift (top left), $g-r$ color (top right), Petrosian half-light radius (derived using $r$-band photometry, bottom left), and the $r$-band axis ratio ($b/a$, bottom right) of the galaxies in the new SwiM catalog, SwiM\_v4.1, as compared to the MaNGA MPL-11 catalog. All the histograms are normalized by the total number of galaxies in each data set. For each panel, the red solid outline denotes the SwiM\_v4.1 distribution, while the gray shaded region represents the MaNGA sample. All the data presented here are from the NASA-Sloan Atlas. The SwiM\_v4.1 catalog has a similar distribution to the MaNGA sample for each of these properties. Please see Sections~\ref{ssec:mdr1} and \ref{ssec:vlw_corr} for details.}
\label{fig:histprop}
\end{figure*}

\begin{figure*}
\centering
\includegraphics[width=0.50\textwidth]{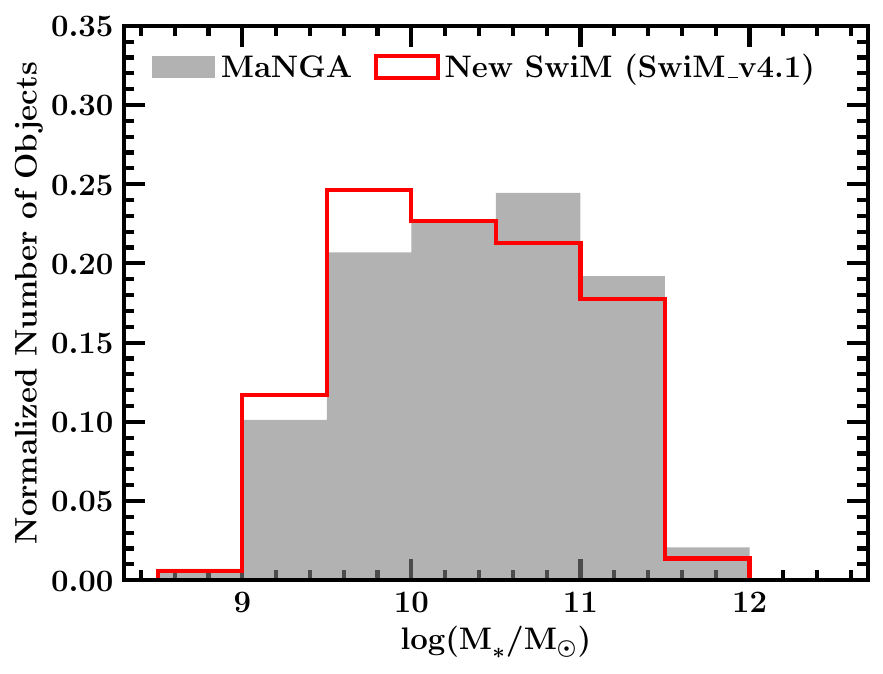}
\includegraphics[width=0.42\textwidth]{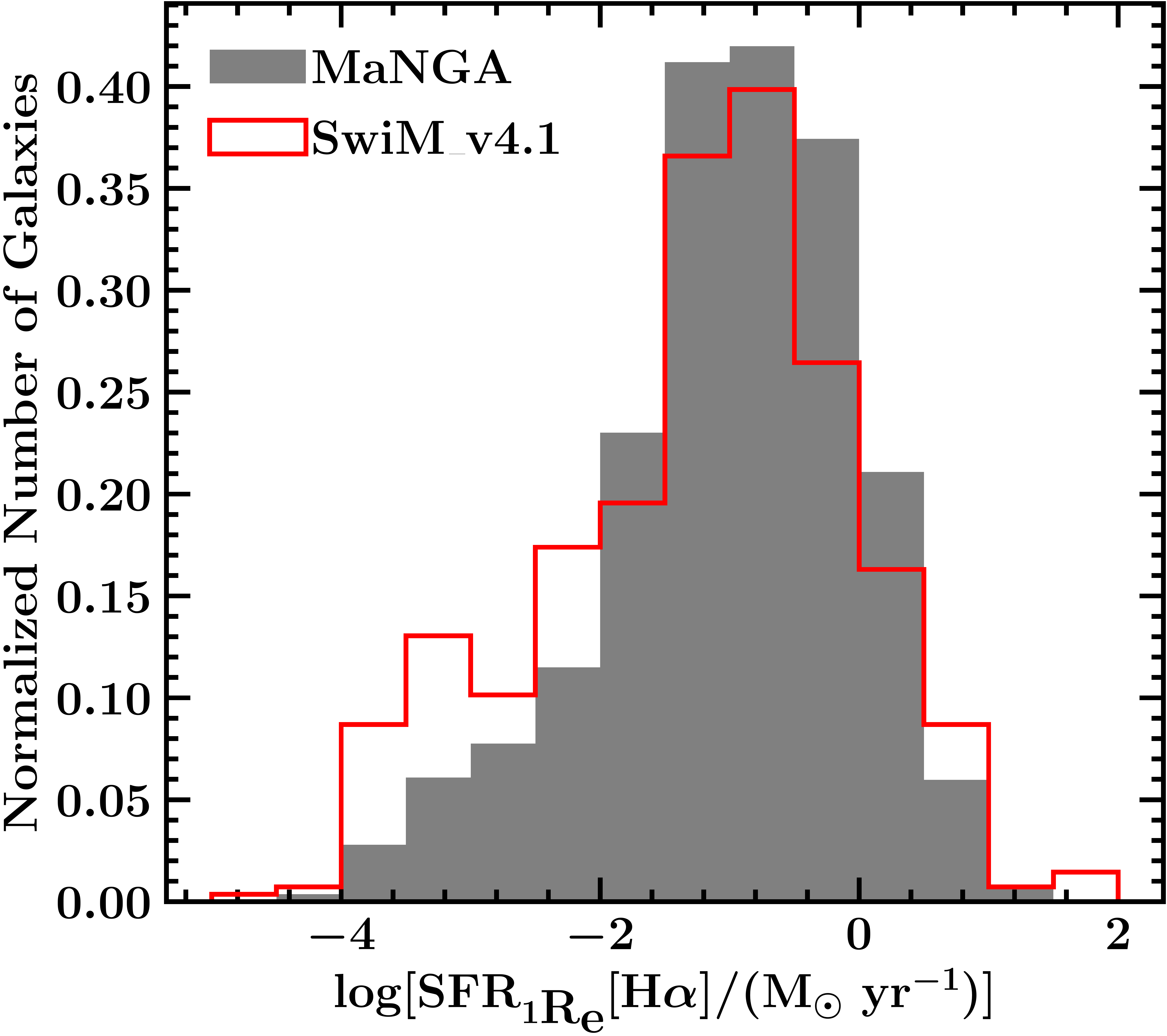}
\includegraphics[width=0.48\textwidth]{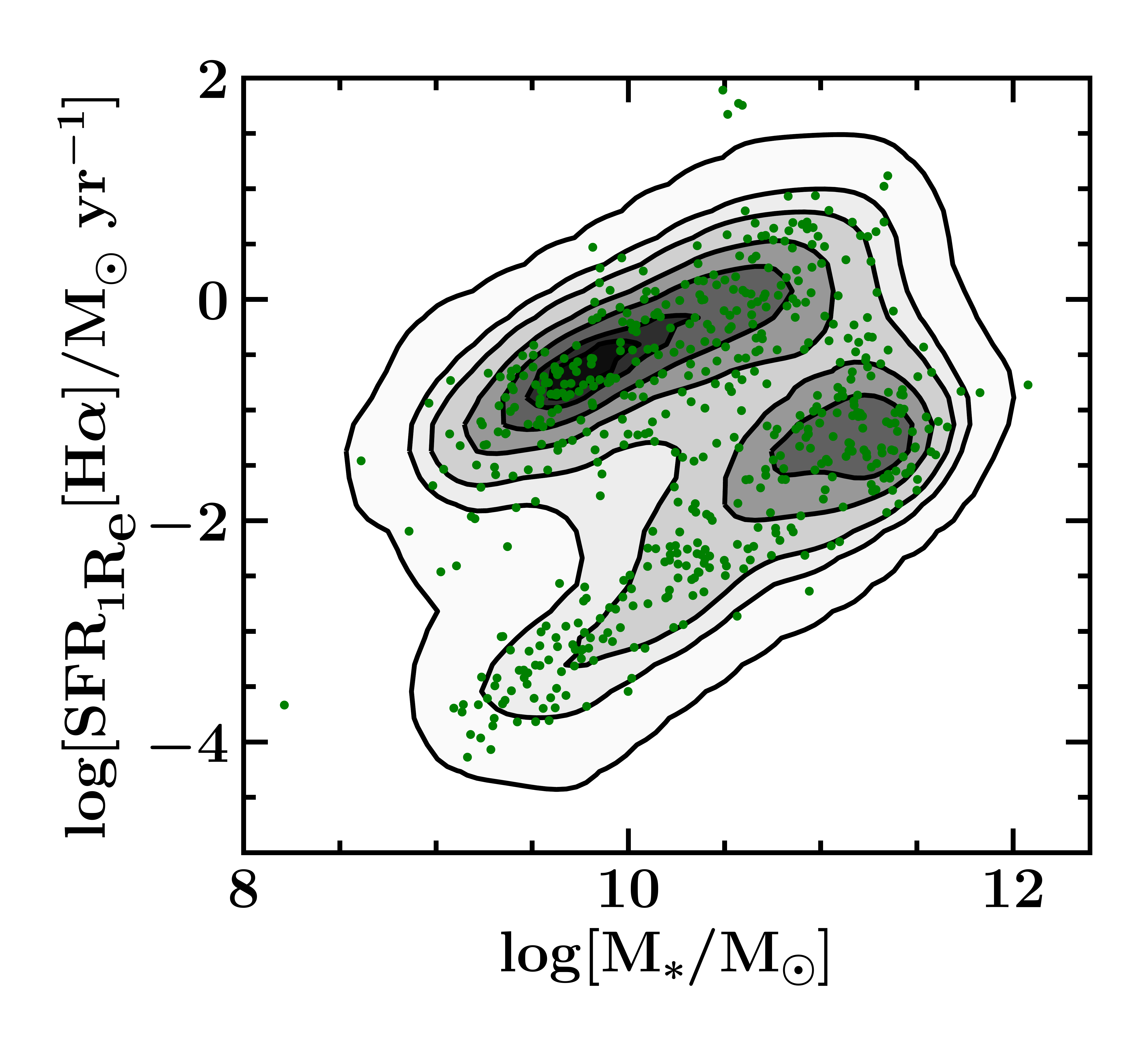}
\caption{\textit{Top Left:} Same as Figure~\ref{fig:histprop}, but for stellar mass. SwiM\_v4.1 is slightly over-dense in low-mass galaxies as compared to MaNGA\null. \textit{Top Right:} Same as Figure~\ref{fig:histprop}, but for the SFR as measured by the total H$\alpha$ luminosity within 1 effective radius ($R_{\rm e}$). The $L$(H$\alpha$) measurements are corrected for both foreground extinction and internal attenuation, assuming the \citet{Odonnell1994} law, R$_V=3.1$, and an intrinsic Balmer ratio of 2.86. SwiM has a significant over-abundance in low-SFR objects. \textit{Bottom:} SFR(H$\alpha$) within $1R_{\rm e}$ vs.~stellar mass for the SwiM\_v4.1 catalog (green filled circles) as compared to the MaNGA MPL-11 catalog (black and grey contours). The SFRs are corrected for reddening as described above. While SwiM\_v4.1 generally recovers both the star-forming and passive galaxy sequences, the catalog is slightly over-populated at the lower-mass, low-SFR end of the diagram and under-populated in the transition region between red and blue galaxies, compared to the MaNGA sample. See Sections~\ref{ssec:mdr1} and \ref{ssec:vlw_corr} for details.}
\label{fig:sfms_data}
\end{figure*}

\subsection{Comparison of SwiM\_v3.1 and SwiM\_v4.1 to MaNGA}\label{ssec:mdr1}
The original SwiM catalog (SwiM\_v3.1) was created using the MPL-7 catalog which had a total of 4706 galaxies. In contrast, SwiM\_v4.1 draws from the MPL-11 catalog which has 10,145 galaxies. Meanwhile, the \swift\ archive has also grown in size at a faster rate than the MaNGA survey between the MPL-7 and MPL-11 data releases. Given both the large difference in the parent MaNGA samples and the growth of the \swift\ archive, we do not directly compare SwiM\_v4.1 and SwiM\_v3.1. Instead, we discuss how each SwiM release compares to its associated MaNGA release. We briefly summarize the comparison of the SwiM\_v3.1 catalog to the MPL-7 version of the MaNGA survey in Section~\ref{ssec:dr1}; a more in-depth comparison is provided in \cite{swimdr1}.
\subsubsection{SwiM\_v4.1 and MPL-11 Version of MaNGA}
While there are many configurations of the ``MaNGA sample,''\footnote{For more information see: \url{https://www.sdss4.org/dr17/manga/manga-target-selection/}} we cross-referenced {\it all} unique galaxies in the MaNGA catalog, including objects observed in ancillary programs. As a result, the final catalog of MaNGA galaxies that we cross-matched with the \swift\ archive has slightly different properties than the pre-defined MaNGA Primary, Primary+, and Full Secondary samples. We discuss the differences in these samples in detail in Section~\ref{ssec:vlw_corr}. Here we compare all unique galaxies in MaNGA MPL-11 to the SwiM\_v4.1 catalog. 

Figure~\ref{fig:histprop} shows the redshift, $g-r$ color, size and axial ratio distributions for the SwiM\_v4.1 and MaNGA galaxies. All quantities are taken from the NASA-Sloan Atlas \citep[NSA;][]{Blanton2011}, and size is defined as the elliptical Petrosian half-light semi-major axis measured for the $r$-band ($R_{50}$).  A majority of the objects in our catalog are nearby and relatively face-on; 78\% the galaxies have $z\lesssim0.05$, 97\% have $R_{\rm Pet,50} < 20^{\prime\prime}$, and 77\% have $b/a \gtrsim 0.6$. The SwiM\_v4.1 distributions presented in Figure~\ref{fig:histprop} are qualitatively similar to those of the MaNGA MPL--11 catalog. In fact, we performed a two-sample Kolmogorov-Smirnov (K-S) test, and found that both the $g-r$ and $R_{50}$ distributions have p-values $p>0.05$, meaning the SwiM\_v4.1 and MaNGA distributions for these properties are not distinct at the 95\% significance level. 

Figure~\ref{fig:sfms_data} presents the distributions of stellar mass, reddening and attenuation-corrected SFR, as described in Section~\ref{ssec:swim_cat}, and the ``star-forming main sequence'' for SwiM\_v4.1 and the MaNGA MPL-11 catalog. Approximately 87\% of the SwiM\_v4.1 catalog galaxies have $\log({\rm M}_*/{\rm M}_\odot)\gtrsim 9.5$. The SwiM\_v4.1 catalog recovers both the star-forming and passive galaxy sequences seen in the MaNGA MPL-11 catalog, but is slightly over-populated at the lower-mass blue end of the diagram and under-populated in the ``green valley'', i.e., the transition region between the star-forming and red passive galaxy sequences. A more thorough discussion of these trends are provided in Section~\ref{ssec:vlw_corr}.

\subsubsection{Comparison of SwiM\_v4.1 and SwiM\_v3.1}\label{ssec:dr1}
While SwiM\_v4.1 is $\sim4$ times larger than SwiM\_v3.1, they are somewhat similar in general properties; a majority of the galaxies in both catalogs have $z\lesssim0.05$, $R_{\rm Pet,50} < 20^{\prime\prime}$, $b/a \gtrsim 0.6$ and $\log({\rm M}_*/{\rm M}_\odot)\gtrsim 9.5$. However, when compared to their respective MaNGA catalogs (MPL-7 for SwiM\_v3.1 and MPL-11 for SwiM\_v4.1), the differences between the two samples are more distinct. The larger sample size of SwiM\_v4.1 also corresponds to an increase in the fraction of the MaNGA sample in the SwiM catalog; SwiM\_v3.1 comprised $\sim3.2$\% of the MaNGA MPL-7 sample, while SwiM\_v4.1 contains $\sim5.5$\% of MaNGA MPL-11. When the SwiM catalogs are compared to their respective MaNGA releases, we find that SwiM\_v4.1 qualitatively looks more similar to the MPL-11 version of MaNGA than SwiM\_v3.1 does to MaNGA MPL-7. We quantify this by creating 2D ratio histograms and performing K-S simulations, which are discussed in more detail in Section~2.4 of \cite{swimdr1} and Section~\ref{ssec:vlw_corr} of this paper, respectively. Finally, the fraction of AGNs identified doubled between SwiM\_v3.1 and SwiM\_v4.1, with $f_{\rm AGN}=0.08$ and 0.16, respectively. A more thorough discussion of our AGN detection techniques are provided in Section~6 of \cite{swimdr1} and Section~\ref{sec:AGN} of this work. 

\subsection{Volume-Limited Weight Correction Factors}\label{ssec:vlw_corr}
The MaNGA sample was not selected to be either magnitude- or volume-limited; instead weight corrections were provided by \citet{Wake2017} to statistically correct each of the different MaNGA sample configurations to a volume-limited data set. We rely on the ``\texttt{ESWEIGHT}'' volume-limited weight corrections \citep[see][for details]{Wake2017}, which includes the Primary, Color-Enhanced, and full Secondary samples. The combination of these three samples is defined as the MaNGA ``main sample''. The Primary sample is designed such that 80\% of the galaxies in the Primary sample are covered by the MaNGA IFU out to 1.5~$R_e$. Meanwhile, the full Secondary sample is defined such that 80\% of the galaxies have coverage out to 2.5~$R_e$. Both the Primary and full Secondary samples are further constrained to have a flat number density distribution in the absolute $i$-band magnitude, and comprise 47\% and 37\% of the MaNGA main sample, respectively. Finally, the Color-Enhanced supplement sample fills in the under-sampled regions of the $NUV-i$ vs.~$M_i$ plane, and includes low-luminosity red galaxies, high-luminosity blue galaxies and ``green valley'' objects. These additional systems comprise 16\% of the main MaNGA sample. Different configurations and their weights are described in detail in \citet{Wake2017}, while the final target selection criteria can be found on the SDSS-IV DR17 website\footnote{\url{https://www.sdss4.org/dr17/manga/manga-target-selection/}}. In addition to the MaNGA main sample, a number of ancillary programs were completed by the MaNGA survey, which produced objects that had MaNGA observations but did not have ``\texttt{ESWEIGHT}'' corrections. Out of the 559 objects in the SwiM\_v4.1 catalog, 47 were from these ancillary programs, and thus are excluded from the calculations described below. 

\begin{figure}
\includegraphics[width=0.5\textwidth]{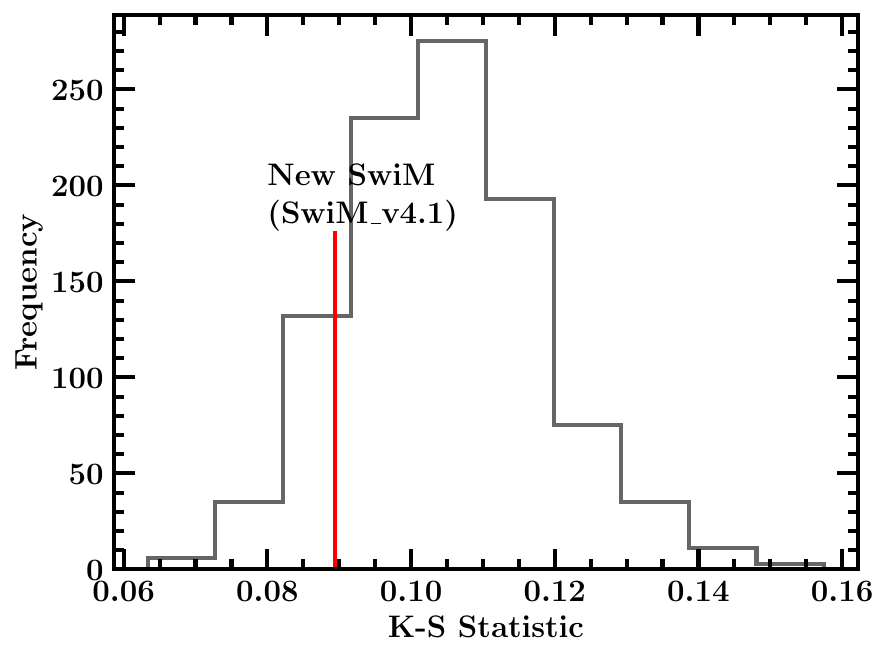}
\caption{Histogram of the K-S test statistic for 1000 random samples drawn from the MaNGA main sample. The test statistic for the SwiM\_v4.1 catalog is denoted by the red vertical line. SwiM\_v4.1 lies at the 13.6 percentile of the distribution, and thus there is only a small statistical probability that a random sample pulled from the MaNGA main sample would have properties similar to the catalog.}\label{fig:rand_samp}
\end{figure}

If the SwiM\_v4.1 catalog is consistent with a random sampling of the MaNGA main sample, then the ``\texttt{ESWEIGHT}'' corrections would allow us to scale our new catalog to a volume-limited galaxy sample.  To test this possibility, we compared the SwiM\_v4.1 data set to randomly selected sets of objects drawn from the MaNGA main sample in the $g-r$ vs.~${\rm M}_*$ plane using a methodology similar to that presented in \cite{swimdr1}. Specifically, we created 1000 samples of 512 galaxies randomly drawn from the MaNGA main sample, and computed the K-S test statistic between the random data sets and SwiM\_v4.1.  The distribution of these values is presented in Figure~\ref{fig:rand_samp}. SwiM\_v4.1 lies at the 13.6 percentile of the distribution, and is thus unlikely to be a true random sample. 

\begin{figure*}
\centering
\includegraphics[width=0.48\textwidth]{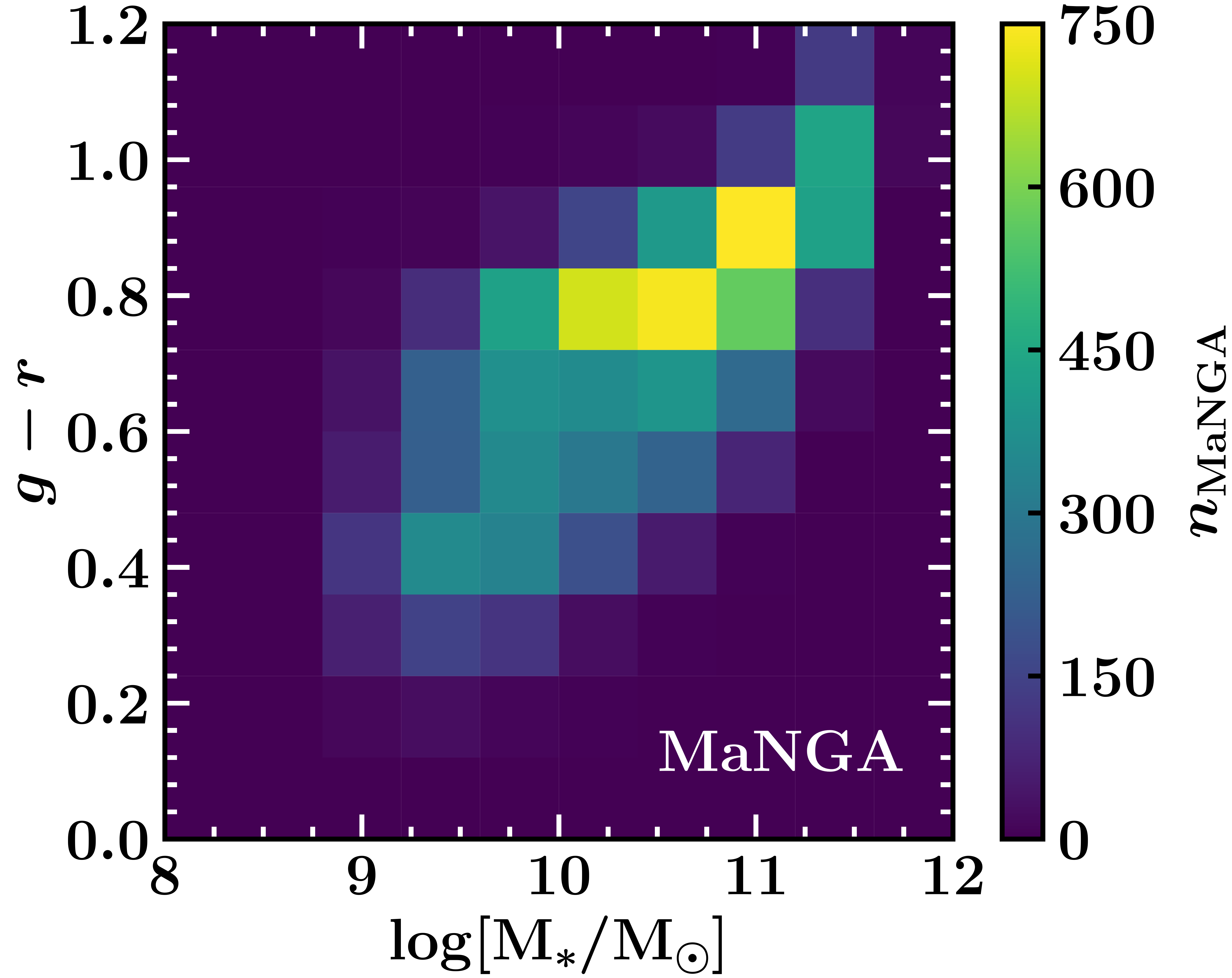}
\includegraphics[width=0.48\textwidth]{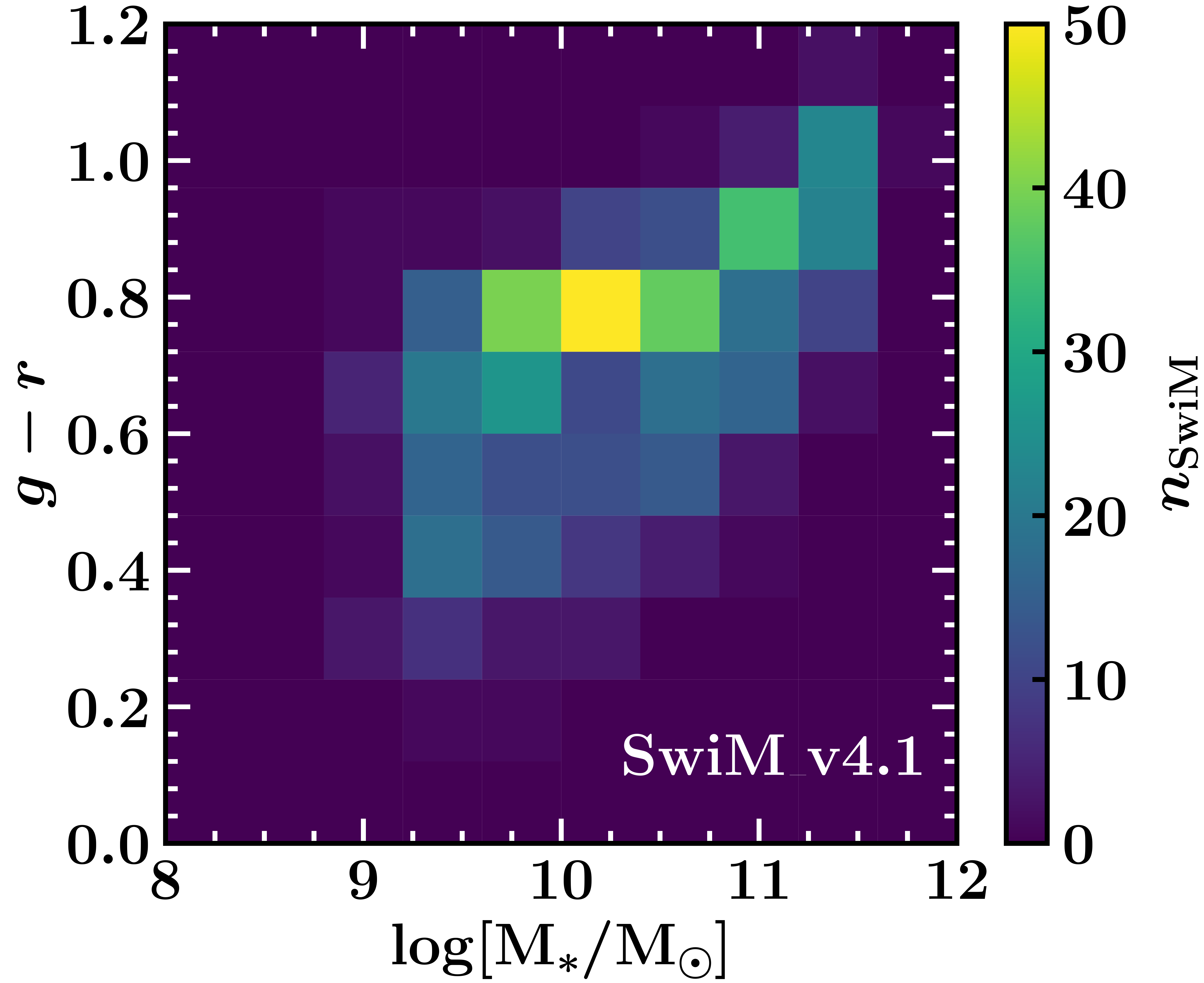}
\includegraphics[width=0.48\textwidth]{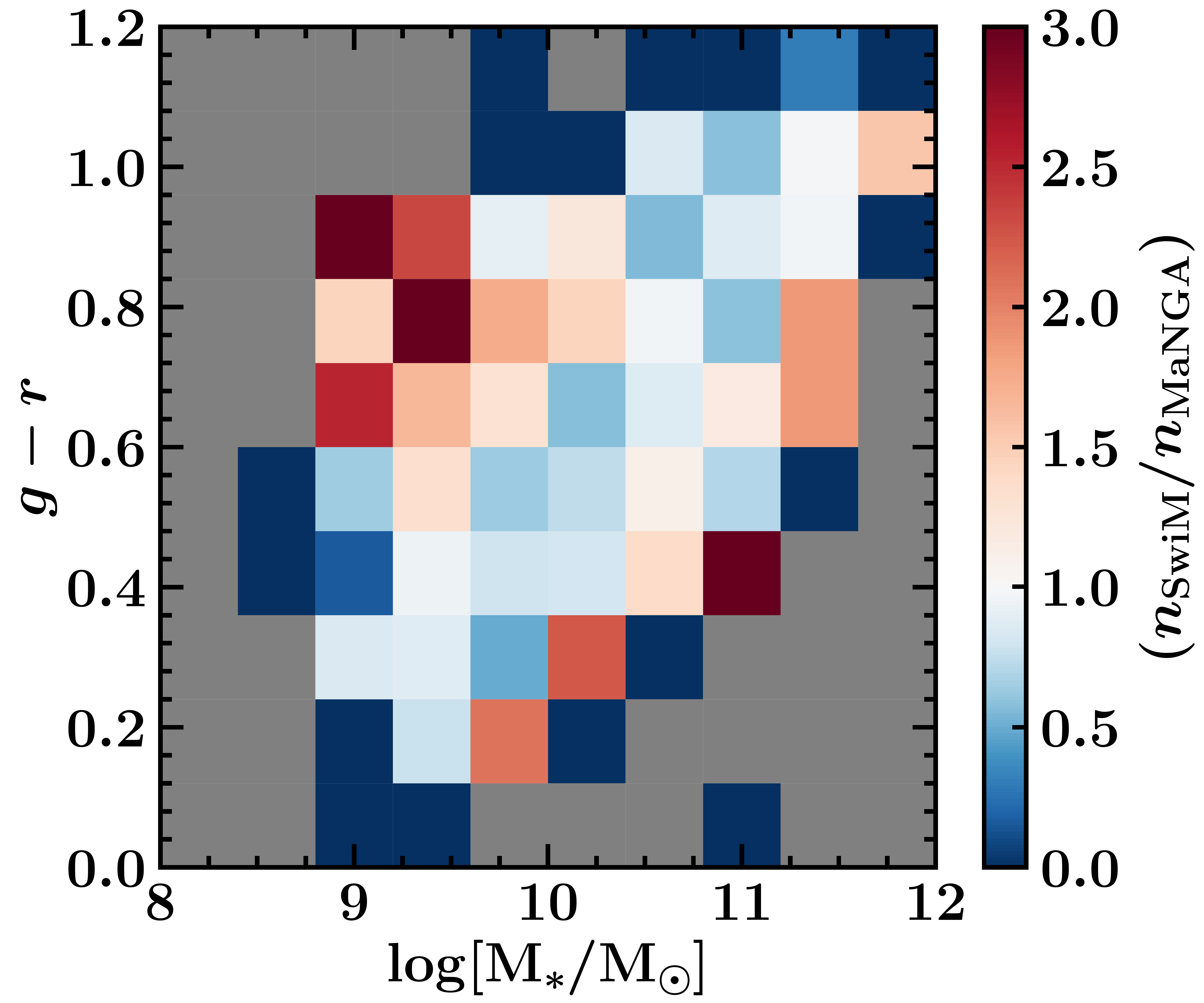}
\caption{2--D histograms of the number distribution of the MPL-11 MaNGA main sample (top left), the SwiM\_v4.1 (top right), and the ratio between the two (bottom center), in $g-r$ vs.~stellar mass. The two sample distributions on the top show high density as yellow, and low density as shades of purple, denoted by the color bar. The MaNGA main sample has a strong peak in the high-mass portion of the red sequence. While the SwiM\_v4.1 catalog appears to have a qualitatively similar distribution, the difference between the two samples is clear in the ratio of the number densities ($n_{\rm{SwiM}}$ and $n_{\rm{MaNGA}}$, respectively) in the bottom panel. Both number densities are calculated by normalizing to the total number of objects in each sample. If the number densities are equal, the bin color is white, while over-densities in SwiM\_v4.1 catalog are represented by shades of red and under-densities by shades of blue, as denoted by the color bar. The inner region red sequence is slightly under-sampled in SwiM\_v4.1, while the lower-mass, red-end of the distribution is over-sampled.}\label{fig:2d_ratio}
\end{figure*}

We then followed the same methodology as \cite{swimdr1} and calculated the scaling corrections needed to match the data sets.  First, we binned the SwiM\_v4.1 and MaNGA main samples into $10\times10$ linearly-spaced intervals in the $g-r$ vs. ${\rm M}_*$ plane, as shown in Figure~\ref{fig:2d_ratio}. Based on the two top panels, the SwiM\_v4.1 catalog appears on average to be redder and higher-mass compared to the MaNGA sample. We divided the 2--D histograms, both of which are normalized by their respective sample sizes, to create the density ratio plot shown in the bottom panel of Figure~\ref{fig:2d_ratio}. The bottom panel illustrates a clear qualitative trend where the inner region red sequence is under-sampled, and the lower-mass, red-end of the distribution is over-sampled. We used the non-normalized, binned density ratio distribution to provide the scaling factors to the ``\texttt{ESWEIGHT}'' values. These scaling factors and their uncertainties are provided in the Swim\_all catalog file. We do note that there are a significant number of bins that the SwiM\_v4.1 catalog does not cover, and caution users to only use scaling factors when the galaxies of interest lie within bins populated by SwiM\_v4.1.

\subsection{Properties of the \swift\ Observations in SwiM\_v4.1}\label{ssec:swft_obs}
\begin{figure*}
    \centering
    \includegraphics[width=0.45\textwidth]{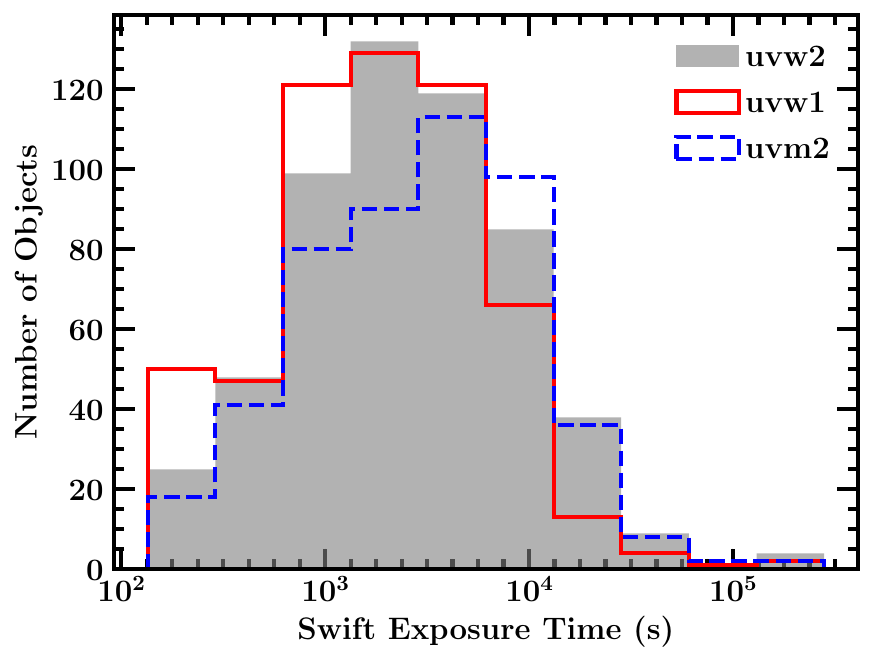}
        \includegraphics[width=0.45\textwidth]{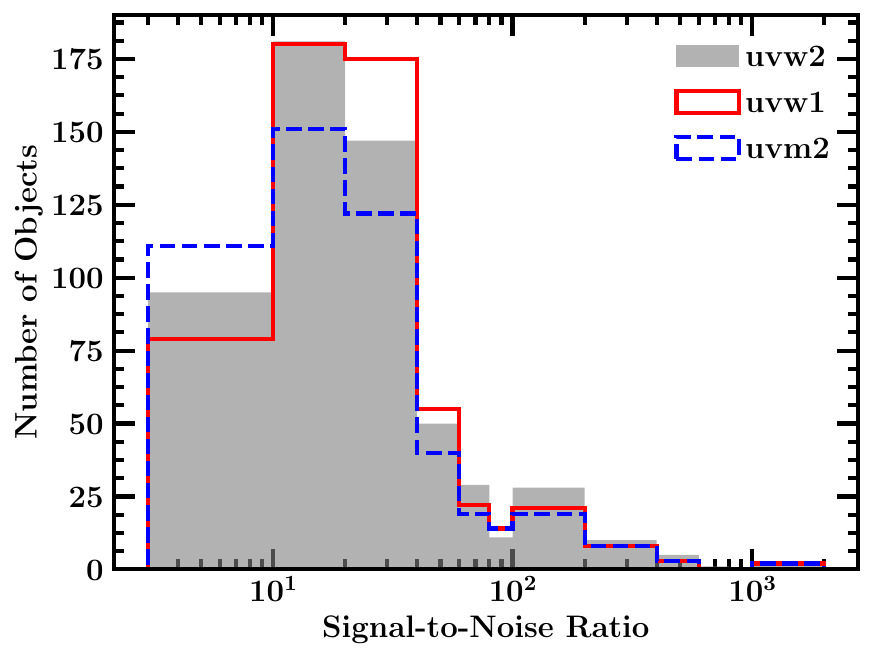}
    \caption{{\it Left: } The distribution of exposure times for each \swift/UVOT filter in seconds, with uvw2 shown as a grey filled histogram, uvw1 as a solid red line and the intermediate-bandpass uvm2 filter displayed as a dashed blue line. {\it Right: }The S/N distribution for the integrated \swift/UVOT NUV fluxes, with a same color scheme as the left panel. We note that $\sim20$\% of galaxies in all three filters have an integrated flux with a ${\rm S/N} < 10$. However, we purposely make no cuts on S/N, so as not to bias the dataset towards particular science questions. A more in-depth discussion on these statistics is provided in Section~\ref{ssec:swft_obs}.}
    \label{fig:swift_obs}
\end{figure*}

The criteria for inclusion in the SwiM\_v4.1 catalog included the availability of uvw1 and uvw2 observations of at least 100~s in duration with no foreground star contamination. As no further quality cuts were made, there are objects in the SwiM catalog that have relatively low S/N \swift\ images. We quantify this by providing the distributions of the \swift\ exposure times and the S/N of the  integrated galaxy flux measurements in Figure~\ref{fig:swift_obs}. The integrated flux calculations are described in Section~\ref{ssec:swim_cat} above.

A majority of the observations in all three \swift\ NUV filters are less than 5~ks, with the median exposure time provided in Table~\ref{table:sw_spec}. As a result, the \swift\ integrated flux measurements for most of the galaxies have a relatively low S/N\null. In fact, $\sim20$\% of the galaxies in our sample have ${\rm S/N} < 10$ in at least one of the two wide-bandpass \swift\ NUV filters (uvw1 and uvw2). We provide more detailed statistics of the S/N for each filter in Table~\ref{table:sw_stat}. While there is a wide range in the S/N between objects within the SwiM\_v4.1 sample, we purposely do not exclude any objects in order to not bias the sample towards any specific type of object. We provide the integrated flux and inverse variance for all three \swift\ filters in the Swim\_all catalog file, which is described in more detail in Section~\ref{ssec:swim_cat} and Appendix~\ref{app:swimcat}.

\begin{deluxetable}{lccc}
  \tablecaption{\textit{Swift}/UVOT NUV Integrated Flux Signal-to-Noise Statistics}
\label{table:sw_stat}
\setlength{\tabcolsep}{10pt}
\tablehead{
{Filter} & {${\rm S/N}<10$} & {${\rm S/N}<50$} & {${\rm S/N}<100$}\\
(1) & (2) & (3) & (4)}
\startdata
uvw2 & \phantom{1}94 (17\%) & 446 (80\%) & 513 (92\%)\\
uvm2 & 110 (23\%)& 404 (83\%)& 457 (94\%)\\
uvw1 & \phantom{1}79 (14\%) & 471 (84\%) & 525 (94\%)\\
\enddata
\tablecomments{Column 1: UVOT Filter, Columns 2--4: Number of galaxies whose integrated flux falls within the given S/N cutoffs of 10, 50 and 100 respectively. The percent of total sample is given in parentheses.}
\end{deluxetable}

\subsection{Overlap between SwiM\_v4.1 and Other Galaxy Catalogs}\label{ssec:ocat}
\begin{deluxetable}{llcc}
  \tablecaption{Cross-Listing of SwiM\_v4.1 with other Catalogs\label{table:xrefcat}}
  \setlength{\tabcolsep}{4pt}
\tablehead{
{Catalog}&{Primary Selection} &{Number of} & {\% of Total}\\
Name& Criterion & Objects & Sample\\
(1) & (2) & (3) & (4)}
\startdata
{2MFGC} & 2MASS $a/b\geq3$ & 19 & 3.4\\
{CGCG} & {$m_{\rm POSS-I} < 15.7$\tablenotemark{a}} &137 & 24.5\\
{IC}\tablenotemark{b} & $m_{\rm B}<13.2$ & 26 & 4.7\\
{KUG} & UV excess $>$ A stars & 37 & 6.6\\
{MCG} &  $m_{\rm POSS-I} < 15$\tablenotemark{a}& 116 & 20.8\\
{Mrk} &{UV excess} & 11 & 2.0 \\
{NGC} & $m_{\rm B}<13.2$ & 34 & 6.1\\
PGC\tablenotemark{c} & Galaxies within $z\approx0.2$ & 201 & 36.0 \\
UGC & Diameter of galaxy $>1$\arcmin & 54 & 9.7 \\
\enddata
\tablecomments{Column 1: Name of other galaxy catalog. Column 2: The main selection criterion of the catalog. Column 3: Number of objects in SwiM\_v4.1 in that catalog. Column 4: Percentage of the total SwiM\_v4.1 sample that has an overlap with the given catalog.}\vspace{-2mm}
\tablenotetext{a}{The CGCG and MCG relied on data from the Palomar Observatory Sky Survey (POSS-I), which used two Kodak red and blue filters (see \small{\url{http://gsss.stsci.edu/SkySurveys/Surveys.htm}}).}\vspace{-2mm}
\tablenotetext{b}{The IC is a supplement to the NGC and as such has the same primary selection criterion.}\vspace{-2mm}
\tablenotetext{c}{The PGC is based on the Lyon-Meudon Extragalactic Database \citep[LEDA;][]{ledapaper}.}
\end{deluxetable}

While SwiM\_v4.1 is by definition a subset of the MaNGA survey, our catalog also has significant overlap with other well-known data sets. In order to determine the overlap, we searched for cross-listings for all the galaxies in SwiM\_v4.1 using the \cite{NED}\footnote{The NASA/IPAC Extragalactic Database (NED) is funded by the National Aeronautics and Space Administration and operated by the California Institute of Technology.} and stored the results in the ``\texttt{NAME}'' column of the SwiM\_all catalog file. Given the large number of galaxy catalogs, we focused our search on the following nine: the Two Micron All-Sky Survey (2MASS) Flat Galaxy Catalog \citep[2MFGC;][]{2MFGCpaper}, the Catalog of Galaxies and of Clusters of Galaxies \citep[CGCG;][]{CGCGpaper}, the Index Catalog \citep[IC;][]{ICpaper}, the Kiso survey for ultraviolet-excess galaxies \citep[KUG;][]{KUGpaper}, the Morphological Catalog of Galaxies \citep[MCG;][]{MGCpaper}, the Markaryan objects \citep[Mrk;][]{Mrkpaper}, the New General Catalog \citep[NGC;][]{NGCpaper}, the Principal Galaxies Catalog \citep[PGC;][]{PGCpaper} and the Uppsala General Catalog of Galaxies \citep[UGC;][]{UGCpaper}. The main selection criterion for each of these catalogs is provided in Table~\ref{table:xrefcat}. If an object was not in any catalog, we recorded NED's preferred object name. Meanwhile, if an object was cross-listed in more than one catalog, we preferentially select the NGC or UGC name, when available.

We present the overlap between SwiM\_v4.1 and each of the nine catalogs listed above in Table~\ref{table:xrefcat}. In total, 237/559, or approximately 42\% of the galaxies in SwiM\_v4.1 were cross-listed in at least one of the nine catalogs described above. Of those 237 objects, 173 or 73\% are cross-listed in at least two catalogs.

\section{Active Galactic Nuclei in SwiM\_\MakeLowercase{v}4.1}\label{sec:AGN}

To identify AGNs, we followed the same steps as for the SwiM\_v3.1 catalog \citep{swimdr1}, which we summarize briefly below.

\begin{figure*}
    \centering
    \includegraphics[width=0.9\textwidth]{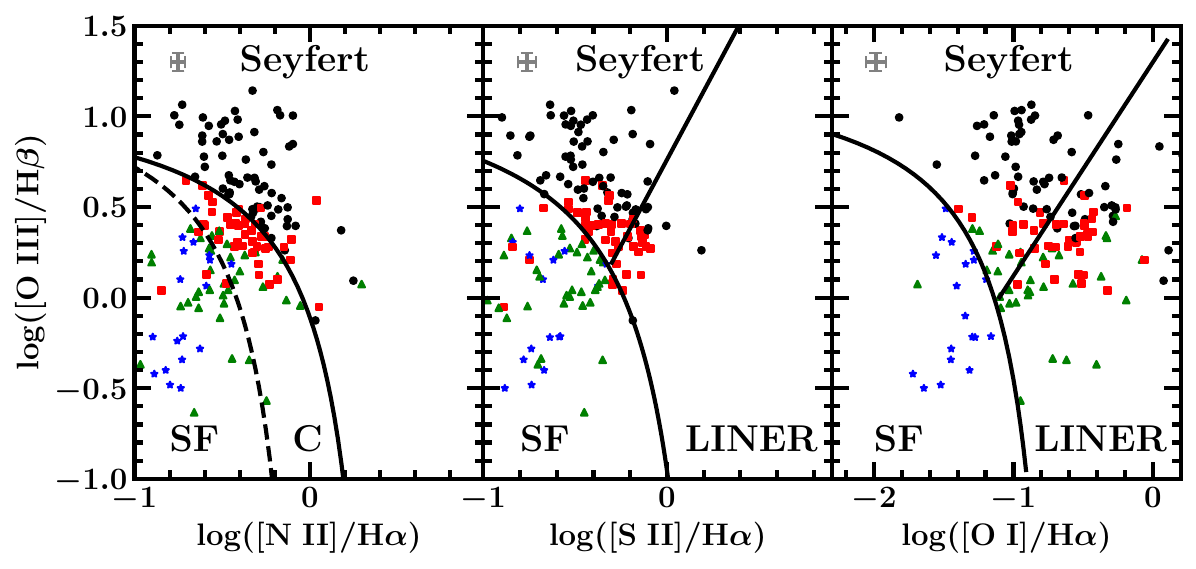}
    \caption{The [\ion{O}{3}]/H$\beta$ vs.~[\ion{N}{2}]/H$\alpha$, [\ion{S}{2}]/H$\alpha$ and [\ion{O}{1}]/H$\alpha$ diagrams for the nuclear resolution element of the AGN candidate galaxies. The extreme starburst, Seyfert, and LINER lines from \citet{Kewley2006} are shown as solid black lines, while the composite line, representing galaxies where emission is dominated neither by star formation nor by AGN activity, from \citet{Kauffmann03} is shown as a dashed line. The different regions within the three diagrams are labelled, with ``SF'' and ``C'' corresponding to the star-forming and composite loci, respectively. The characteristic error bar for each diagram is shown in the upper left corner. Galaxies outside of the star-forming locus on all three diagrams are shown as black circles, those outside of the locus in two of the three diagrams are shown as red squares, those outside in just one are shown as green triangles and those in the star-forming locus in all three diagrams are shown as blue stars. There are 67 galaxies that fall outside the star-forming locus in all three diagrams.}    \label{fig:bpt}
\end{figure*}

We first identified AGN candidates by inspecting spatially resolved emission-line diagnostic diagrams. Following \citet{Kauffmann03} and \citet{Kewley2006}, we employed three diagrams that plot [\ion{O}{3}]/H$\beta$ versus [\ion{S}{2}]/H$\alpha$, [\ion{N}{2}]/H$\alpha$, or [\ion{O}{1}]/H$\alpha$ based on MaNGA spectra and flagged any objects with 10 or more MaNGA pixels within 0.3 $R_e$ that fall in the Seyfert or LINER regions of the diagrams. We added any galaxies that had SDSS classifications of `AGN', `QSO', or `Broadline'. This resulted in 173 initial candidates. We refined the above procedure by measuring the same diagnostic line ratios in a single resolution element centered on the nucleus of each galaxy namely, a circular aperture of diameter $2\farcs92$. This procedure yielded 67 galaxies that likely host AGNs. The resulting emission-line ratio diagrams can be seen in Figure~\ref{fig:bpt}. 

We also identified galaxies with X-ray emission detected by the \textit{Swift} X-ray Telescope (XRT) by cross-referencing our galaxies with the \swift\ Point Source Catalog \citep{Evans2007}. In our sample, a total of 99 galaxies were detected by \textit{Swift}/XRT. We converted the observed 2-10~keV XRT count rates or upper limits to a flux with the help of the {\tt WebPIMMS} simulator\footnote{\url{https://heasarc.gsfc.nasa.gov/cgi-bin/Tools/w3pimms/w3pimms.pl}} and then to a luminosity. In the process we assumed a power-law X-ray spectrum with a photon index of 1.8 and foreground absorption column density appropriate for the line of sight to that galaxy. We note for reference that, under the above assumptions, an XRT count rate of 1~s$^{-1}$ from a galaxy at a distance of 100~Mpc (close to the median of our sample) corresponds to a 2--10~keV luminosity of $L_\mathrm{HX}\sim 1\times 10^{44}~{\rm erg~s}^{-1}$. In comparison, the majority of the galaxies observed with the XRT have count rate upper limits of order $10^{-3}~{\rm s}^{-1}$, which implies upper limits of $L_{\rm HX} < 10^{41}~{\rm erg~s}^{-1}$. We do not report the individual upper limits for galaxies that were not detected by the \textit{Swift}/XRT because these do not have much diagnostic value; nearby Seyfert galaxies can have X-ray luminosities near or below this threshold and nearby starburst galaxies can have X-ray luminosities near or above this threshold.\footnote{For example, the Seyfert galaxy NGC~4151 has $L_{\rm HX}\sim 1\times 10^{42}~{\rm erg~s}^{-1}$ \citep{Nandra2007} and the low-luminosity Seyfert galaxies M81 and M58 have $L_{\rm HX} \sim 2\times 10^{39}~{\rm erg~s}^{-1}$ and $\sim 2\times 10^{41}~{\rm erg~s}^{-1}$, respectively \citep{Verrecchia2007,Eracleous2002}. In contrast, the starburst galaxy NGC~3256 has $L_{\rm HX} \sim 2\times 10^{41}~{\rm erg~s}^{-1}$ \citep{Moran1999}.}

The PSF of the \textit{Swift}/XRT has a half-energy width of $17.\!\!^{\prime\prime}$4--$17.\!\!^{\prime\prime}8$ at 4~keV \citep{Moretti2005}, which encompasses a significant portion of every galaxy in our sample. We thus must consider the potential contribution of X-ray binaries (XRBs) on the X-ray emission we observe. The contribution from low-mass XRBs can be parameterized via a galaxy's total stellar mass, and the contribution from high-mass XRBs is quantified using a galaxy's star formation rate \citep[e.g.,][]{Fabbiano2006, Lehmer2010}. Therefore to classify an object as an X-ray AGN, we applied Equation~(3) of \cite{Lehmer2010} and required that the observed 2--10~keV X-ray luminosity ($L_\mathrm{HX}^{\mathrm{gal}}$) be higher than the maximum possible contribution from the XRB population.  Of the 99 galaxies with X-ray detections, 37 were luminous enough to be regarded as AGNs.

In summary, we detected 93 AGNs through the combination of the above tests. Of these, 67 are galaxies whose nuclear relative emission line strengths place them in the AGN region of the narrow-line diagnostic diagrams. Out of those 67 galaxies, 11 also have X-ray luminosities at least one order of magnitude higher than one would expect from XRBs based on their stellar masses and star formation rates. Moreover, 26 additional galaxies are identified as AGNs solely by their X-ray luminosities. The MaNGA IDs, names, and details of their identification can be found in Table \ref{table:agn_candidates}, of which only a portion is presented here. The full table can be found on the electronic version of the {\it Astrophysical Journal}. We note that we likely have not identified all of the AGNs in SwiM\_v4.1. For example, we likely missed any AGNs that were too faint to be detected by our detection techniques or hidden by host-galaxy star-formation \citep[e.g.,][]{Moran2002}. While a more thorough approach employing more detection techniques could help identify more AGNs within the catalog, such analysis is beyond the scope of this work.

We also compared our final list of SwiM\_v4.1 AGNs with AGNs identified in the MaNGA catalog \citep{Comerford2020}. We used an updated version of this catalog that includes 387 AGNs out of the final sample of 10,018 MaNGA galaxies; this list was kindly made available to us by J. Comerford (Comerford et al., in prep). Their AGNs were selected among MaNGA galaxies using four criteria unrelated to emission line diagnostic diagrams: (a) WISE mid-infrared color cuts, (b) {\it Swift}/BAT hard (14--195~keV) X-ray detections, (c) 1.4~GHz radio luminosity relative to the galaxy stellar mass and H$\alpha$ luninosity, and (d) presence of broad emission lines in the MaNGA nuclear spectra. Of our 93 AGNs, 26 are also included in the Comerford et al.\ catalog.

\begin{deluxetable}{cccc}
  \tablecaption{X-ray and Narrow-Line Diagnostic Diagram Classifications of AGN Candidates}\label{table:agn_candidates}
  \setlength{\tabcolsep}{10pt}
  \tabletypesize{\footnotesize}
\tablehead{
{}&{$\log(L_{\mathrm{HX}}^{\mathrm{gal}}\;/$}&{$\log(L_\mathrm{HX}\;/$}&{Narrow-Line}\\
{MaNGA ID} & erg s$^{-1}$) & erg s$^{-1}$) & {Classification}\\
{(1)} & (2) & (3) & (4)}
\startdata
1-120103&\dots&\dots&S/S/S\\ 
1-121604&\dots&\dots&S/L/L\\
1-135059&\dots&\dots&S/S/L \\
1-137883&38.6&39.8&S/S/S\\
1-150842&\dots&\dots&S/L/L\\
\enddata
\tablecomments{Column 1: MaNGA ID, Column 2: Maximum contribution from XRBs according to Equation (3) of \cite{Lehmer2010}, Column 3: Observed hard X-ray luminosity (2--10~keV) from \textit{Swift}/XRT, Column 4: Classification in the narrow-line diagnostic diagrams using the criteria from \cite{Kewley01}, \cite{Kauffmann03} and \cite{Kewley2006}. The labeling refers to the galaxy's location in the [\ion{O}{3}]/H$\beta$ vs.~[\ion{N}{2}]/H$\alpha$, [\ion{S}{2}]/H$\alpha$ and [\ion{O}{1}]/H$\alpha$ diagrams, respectively. H: star-forming locus S: Seyfert, C: Composite region defined by \cite{Kauffmann03} in the [\ion{O}{3}]/H$\beta$ vs.~[\ion{N}{2}]/H$\alpha$ diagram, L: LINER. The entirety of Table \ref{table:agn_candidates} is published in the electronic edition of the {\it Astrophysical Journal}. We show a portion here to provide information on its form and content.}
\end{deluxetable}

\section{Summary}\label{sec:summ}
In this work, we present the second data release of the Swift+MaNGA, or SwiM catalog. The matching pixel scale and angular resolution of the NUV and optical data used to construct SwiM\_v4.1 make it an ideal catalog to study both the progression of star-formation quenching and characterize the dust attenuation laws in nearby galaxies. The updated sample has a total of 559 galaxies that have both MaNGA and \swift/UVOT uvw1 and uvw2 observations; this is about 4 times larger in size than the first release. We provide the same data as in the original release, including integrated \swift/UVOT photometry and scaling factors needed to translate the catalog into a volume-limited sample, as well as new data, such as additional emission-line maps and a separate file that reports the percentage of science-ready pixels in each MaNGA map. All of the images and MaNGA maps have been transformed to the same wcs, resolution, and pixel-sampling as the \swift\ uvw2 images,  with a final resolution of $2\farcs9$ and a 1\arcsec\ per pixel scale. This uniformity across both the maps of\ an individual galaxy and over all the galaxies in the sample allow for seamless comparisons of different physical measurements across the faces of galaxies. Finally, we use a combination of optical emission-line ratios and X-ray observations to identify AGNs within the SwiM\_v4.1 sample. 

This dataset thus provides a unique look into the evolution of individual galaxies and nearby galaxies as a whole with unprecedented spatial and wavelength coverage at the same resolution. We make the SwiM\_v4.1 catalog publicly available on the SDSS website as an update to the original SwiM VAC\footnote{A description of the SwiM VAC is provided here: \url{https://www.sdss4.org/dr17/data_access/value-added-catalogs/?vac_id=swift-manga-value-added-catalog}, and the data are stored on the SDSS SAS: \url{https://data.sdss.org/sas/dr17/manga/swim/v4.1/}}.

\section{Acknowledgments}

\noindent We thank Michael Blanton for a critical reading of the manuscript and helpful comments. We also thank Joel Brownstein for his insightful comments and his help in finalizing, staging and releasing the new SwiM catalog.

\noindent This work was supported by NASA through grant numbers 80NSSC20K0436 (ADAP) and the {\it Swift} GI Program ID 16190136. This work was supported by funding from Ford Foundation Postdoctoral Fellowship, administered by the National Academies of Sciences, Engineering, and Medicine, awarded to MM in 2021-2022. The work of MM is supported in part through a fellowship sponsored by the Willard L. Eccles Foundation.

\noindent This work made use of data supplied by the UK Swift Science Data Centre at the University of Leicester. This research has made use of data and/or software provided by the High Energy Astrophysics Science Archive Research Center (HEASARC), which is a service of the Astrophysics Science Division at NASA/GSFC and the High Energy Astrophysics Division of the Smithsonian Astrophysical Observatory. This research made use of Astropy, a community-developed core Python package for Astronomy \citep{Whelan2018}.  This research has made use of the NASA/IPAC Extragalactic Database (NED), which is operated by the Jet Propulsion Laboratory, California Institute of Technology, under contract with the National Aeronautics and Space Administration. The Institute for Gravitation and the Cosmos is supported by the Eberly College of Science and the Office of the Senior Vice President for Research at the Pennsylvania State University. 

\noindent Funding for the Sloan Digital Sky Survey IV has been provided by the Alfred P. Sloan Foundation, the U.S. Department of Energy Office of Science, and the Participating Institutions. SDSS-IV acknowledges support and resources from the Center for High-Performance Computing at the University of Utah. The SDSS web site is www.sdss.org. 
SDSS-IV is managed by the Astrophysical Research Consortium for the Participating Institutions of the SDSS Collaboration including the Brazilian Participation Group, the Carnegie Institution for Science, Carnegie Mellon University, the Chilean Participation Group, the French Participation Group, Harvard-Smithsonian Center for Astrophysics, Instituto de Astrof\'isica de Canarias, The Johns Hopkins University, Kavli Institute for the Physics and Mathematics of the Universe (IPMU) / University of Tokyo, the Korean Participation Group, Lawrence Berkeley National Laboratory, Leibniz Institut f\"ur Astrophysik Potsdam (AIP),  Max-Planck-Institut f\"ur Astronomie (MPIA Heidelberg), Max-Planck-Institut f\"ur Astrophysik (MPA Garching), Max-Planck-Institut f\"ur Extraterrestrische Physik (MPE), National Astronomical Observatories of China, New Mexico State University, New York University, University of Notre Dame, Observat\'ario Nacional / MCTI, The Ohio State University, Pennsylvania State University, Shanghai Astronomical Observatory, United Kingdom Participation Group,Universidad Nacional Aut\'onoma de M\'exico, University of Arizona, University of Colorado Boulder, University of Oxford, University of Portsmouth, University of Utah, University of Virginia, University of Washington, University of Wisconsin, Vanderbilt University, and Yale University.

%

\vspace{5mm}
\facilities{Swift(XRT and UVOT), SDSS-IV/MaNGA}


\software{astropy \citep{Whelan2018},
          Source Extractor \citep{Bertin1996},
          Swift-UVOT-Pipeline (\url{https://github.com/malmolina/Swift-UVOT-Pipeline})
          }



\appendix

\section{SwiM\_\MakeLowercase{v}4.1 catalog Data Model}\label{app:swimcat}
\begin{deluxetable}{lrl}[h]
  \tablecaption{SwiM\_v4.1 Catalog Data Model} \label{table:catalog_dm}
\renewcommand{\arraystretch}{1.1}
\tablehead{
{Column} & {Units} & {Description}}
\startdata
{\texttt{MANGAID}} & \dots & MaNGA ID for the object (e.g., 1-109679)\\
\texttt{PLATE} & \dots & Plate ID for the object\\
\texttt{IFUDSGN} & \dots & IFU design ID for the object (e.g., 12701)\\
\texttt{MNGTARG1} & \dots & MANGA\_TARGET1 maskbit for the galaxy target catalog\\
\texttt{MNGTARG3} &\dots & MANGA\_TARGET3 maskbit for the galaxy target catalog\\
\texttt{DRP3QUAL} & \dots & Quality bitmask from drpall file\\
\texttt{NAME} & \dots & Galaxy Name\\
\texttt{SDSS\_CLASS} & \dots & SDSS DR17 object classification\\
\texttt{EBV} & \dots & $E(B-V)$ value from the \cite{Schlegel1998} dust map \\
\texttt{RA} & {deg} & Right-ascension of the galaxy center in J2000\\
\texttt{DEC} & {deg} & Declination of the galaxy center in J2000\\
\texttt{NSA\_ELPETRO\_PHI} & {deg} & Position angle (east of north) used for elliptical apertures\\
\texttt{NSA\_ELPETRO\_TH50\_R} & {arcsec} & Elliptical Petrosian 50\% enclosed light radius (semi-major axis) in SDSS $r$-band\\
\texttt{NSA\_ELPETRO\_TH50} & {arcsec} & Azimuthally averaged SDSS-style Petrosian 50\% light radius derived from SDSS $r$-band\\
\texttt{NSA\_ELPETRO\_BA} & \dots & Axis ratio used for elliptical apertures\\
\texttt{NSA\_ELPETRO\_MASS} & $h^{-2}$ solar masses & Stellar mass from K-corrected fit for elliptical Petrosian fluxes\\
\texttt{NSA\_Z} & \dots & Heliocentric redshift from the NASA-Sloan Atlas\\
\multirow{2}{*}{}\texttt{NSA\_ELPETRO\_FLUX} & nanomaggies & Elliptical SDSS-style Petrosian flux in the \galex\ and SDSS [FN$u$$g$$r$$i$$z$]\\ & & filter bands (using the $r$-band aperture)\\
\texttt{NSA\_ELPETRO\_FLUX\_IVAR} & nanomaggies$^{-2}$ & Inverse variance of \texttt{NSA\_ELPETRO\_FLUX} [FN$u$$g$$r$$i$$z$]\\
\multirow{2}{*}{}\texttt{SWIFT\_ELPETRO\_FLUX} & nanomaggies & Elliptical SDSS-style Petrosian flux in bands [uvw2, uvm2, uvw1]\\ & & (aperture corrected using r-band aperture)\\
\texttt{SWIFT\_ELPETRO\_FLUX\_IVAR} & nanomaggies$^{-2}$ & Inverse variance for \multirow{2}{*}{}\texttt{SWIFT\_ELPETRO\_FLUX} [uvw2, uvm2, uvw1]; \\ & & if there is no uvm2 measurement the element is $-999$\\
\multirow{3}{*}{}\texttt{SWIFT\_EXPOSURE} & sec & Exposure times for Swift/UVOT bands [uvw2, uvm2, uvw1]; \\ & & if there is no uvm2 measurement the element is $-999$\\
\multirow{3}{*}{}\texttt{APERCORR} & \dots & Aperture correction factor $f_a/f_b$ for Swift/UVOT bands [uvw2, uvm2, uvw1];\\ & & $f_a$ and $f_b$ are the $r$-band integrated fluxes (of the mock galaxy) before and after \\ & & the Swift/UVOT PSF convolution {\citep[see][and Section~\ref{ssec:dr2prod} of this work]{swimdr1}}\\
\texttt{SFR\_1RE} & \dots & Dust corrected log(SFR/$1\;{\rm M}_\odot\;$yr$^{-1}$) using the H$\alpha$ flux within one effective radius reported\\ & & in MaNGA DAP \citep[see][and Section~\ref{ssec:dr2prod} of this work]{Westfall2019,swimdr1}\\
\texttt{SCALING\_FACTOR} & \dots & Scaling factors that represent the number of objects in the SwiM catalog divided by the number \\ & &   in the MaNGA main sample for each 2--D bin {\citep[see][and Section~\ref{ssec:dr2prod} of this work]{swimdr1}}\\
\texttt{SCALING\_FACTOR\_ERR} & \dots & 1$\sigma$ uncertainty for \texttt{SCALING\_FACTOR}\\
\texttt{ESWEIGHT} & \dots & Volume weights from MPL-11 for Primary+ and full Secondary sample \citep[see][]{Wake2017}\\
\enddata
\end{deluxetable}
This appendix provides the SwiM\_v4.1 VAC data model for the SwiM\_all catalog file.  This catalog file holds the basic properties of the galaxies included in the \swim\ catalog, as well as the integrated \galex, \swift/UVOT, and SDSS photometry. The only difference between the original and new SwiM catalog data model is the addition of DRP3QUAL, or the quality bitmask from the MaNGA \texttt{drpall} file. We also note that \texttt{NSA\_ELPETRO\_THETA} has been replaced with \texttt{NSA\_ELPETRO\_TH50} to maintain consistency with the latest version of the MaNGA \texttt{drpall} file. The names and contents of each extension in this file are given in Table~\ref{table:catalog_dm}. All null values in this table have been assigned the value $-999$.

\clearpage
\section{SwiM\_\MakeLowercase{v}4.1 Map HDU Data Models}\label{app:swimmap}
This appendix presents the data model for the SwiM\_v4.1 VAC map files. The map files contain the spatially-matched MaNGA IFU maps, as well as the \swift/UVOT and SDSS photometry. The map data files have 17 total header data units (HDUs), with 5 main groups: D$_n$(4000) in HDU 0, spectral indices (HDU 1--8), emission line fluxes and equivalent widths (HDU 9--14), \swift\ and SDSS photometry (HDU 15--16), and the raw \swift\ data (HDU 17). We describe the HDU format for each group below, including notes on how to use the maps. The names and descriptions of the HDUs are given in Table~\ref{table:maps_dm}, while the formatting of the HDUs are described in Tables~\ref{table:d4000_dm} through~\ref{table:rawswft_dm}. {The SwiM\_v4.1 map files include more emission-line maps than the original catalog, as described in Section~\ref{ssec:dr2prod} of this work.}

All MaNGA maps and UVOT images have masks, where science-ready pixels are indicated by 0, and 1 otherwise. The MaNGA masks are based on those in DR17, but have been simplified to a 0 or 1, given the analysis presented in \cite{swimdr1} and this work. The masks for the UVOT images only affect one object as discussed in Section~\ref{ssec:sw_pipeline}.\\

\begin{deluxetable}{clcrl}[h]
  \tablecaption{SwiM Maps HDU Descriptions\label{table:maps_dm}}
  \renewcommand{\arraystretch}{1.1}
  \tablewidth{0pt}
  \setlength{\tabcolsep}{2pt}
\tablehead{
{Index} &{Name} & {Channels} &{Units} & {Description}}
\startdata
\multirow{2}{*}{} 0 & \texttt{Dn4000} & {5} & {erg s$^{-1}$ cm$^{-2}$ Hz$^{-1}$ arcsec$^{-2}$} & Maps required to calculate the D$_n$(4000) measurements and its uncertainty\\
{1} & \texttt{SPECINDX\_FLUX} & {43} & {erg s$^{-1}$ cm$^{-2}$ arcsec$^{-2}$} & Spectral index flux maps \citep[$F_{I}$ in equation~7 from][]{swimdr1}\\
{2} & \texttt{SPECINDX\_CONT} & {43} & {erg s$^{-1}$ cm$^{-2}$ \AA$^{-1}$ arcsec$^{-2}$} & Spectral index continuum maps \citep[$F_{\rm C0}$ in equation 10 in][]{swimdr1}\\
{3} & \texttt{SPECINDX\_FLUX\_SIGMA} & {43} & erg s$^{-1}$ cm$^{-2}$ arcsec$^{-2}$ & 1$\sigma$ uncertainties for \texttt{SPECINDX\_FLUX}\\
{4} & \texttt{SPECINDX\_CONT\_SIGMA} & {43} & {erg s$^{-1}$ cm$^{-2}$ \AA$^{-1}$ arcsec$^{-2}$} & 1$\sigma$ uncertainties for \texttt{SPECINDX\_CONT}\\
\multirow{2}{*}{}{5} & \texttt{SPECINDX\_MASK} & {43} & \dots & Masks for \texttt{SPECINDX\_FLUX}, \texttt{SPECINDX\_FLUX\_SIGMA}, \texttt{SPECINDX\_CONT} and \\ & & & & \texttt{SPECINDX\_CONT\_SIGMA} \\
{6} & \texttt{COMBINED\_DISP} & {43} & {km s$^{-1}$} & Flux-weighted combined dispersion maps\\
{7} & \texttt{COMBINED\_DISP\_SIGMA} & {43} & {km s$^{-1}$} & 1$\sigma$ uncertainties for \texttt{COMBINED\_DISP}\\
{8} & \texttt{COMBINED\_DIPS\_MASK} & {43} & \dots & Masks for \texttt{COMBINED\_DISP} and \texttt{COMBINED\_DISP\_SIGMA}\\
{9} & \texttt{ELINE\_FLUX} & {22} & {10$^{-17}$ erg s$^{-1}$ cm$^{-2}$ arcsec$^{-2}$} & Gaussian-fitted emission line flux maps based on MPL-11 DAP\\
{10} & \texttt{ELINE\_FLUX\_SIGMA} & {22} & {10$^{-17}$ erg s$^{-1}$ cm$^{-2}$ arcsec$^{-2}$} & 1$\sigma$ uncertainties for \texttt{ELINE\_FLUX}\\
{11} & \texttt{ELINE\_FLUX\_MASK} & {22} & \dots & Masks for \texttt{ELINE\_FLUX} and \texttt{ELINE\_FLUX\_SIGMA}\\
{12} & \texttt{ELINE\_EW} & {22} & {\AA} & Gaussian-fitted equivalent width maps based on MPL-11 DAP\\
{13} & \texttt{ELINE\_EW\_SIGMA} & {22} & {\AA} & 1$\sigma$ uncertainties for the \texttt{ELINE\_EW\_SIGMA}\\
{14} & \texttt{ELINE\_EW\_MASK} & {22} & \dots & Masks for \texttt{ELINE\_EW} and \texttt{ELINE\_EW\_SIGMA}\\
\multirow{2}{*}{}{15} & \texttt{SWIFT/SDSS} & {8} & {nanomaggies} & \textit{Swift}/UVOT and SDSS sky-subtracted images [uvw2,uvw1,uvm2,$u$,$g$,$r$,$i$,$z$]\\
{16} & \texttt{SWIFT/SDSS\_SIGMA} & {8} & {nanomaggies} & 1$\sigma$ uncertainties for \texttt{SWIFT/SDSS}\\
\multirow{2}{*}{}{17} & \texttt{SWIFT\_UVOT} & {12} & \dots &  \swift/UVOT non-sky-subtracted counts, exposure, counts error and mask maps \\ & & & &  [uvw2, uvw1, uvm2]\\
\enddata
\end{deluxetable}

\clearpage
\noindent\textbf{HDU 0: D4000} -- This HDU contains the maps necessary to calculate D$_n$(4000) measurements and their uncertainties. The data are in a 3D array with the third dimension having a size of 5 corresponding to the two data channels, their uncertainties, and the mask. D$_n$(4000) is defined as $\textrm{D}_n(4000) = f_{\nu,\rm{red}}/f_{\nu,\rm{blue}}$, and its uncertainty as
\begin{equation}\label{eqn:d4000_sig}
\sigma_{\textrm{D}_n(4000)}=\textrm{D}_n(4000)\sqrt{\left(\frac{\sigma_{f_{\nu,\rm{red}}}}{{f_{\nu,\rm{red}}}}\right)^{\!\!2}+\left(\frac{\sigma_{f_{\nu,\rm{blue}}}}{{f_{\nu,\rm{blue}}}}\right)^{\!\!2}}. 
\end{equation}
While covariance has been properly accounted for in our data processing, the final uncertainty must be multiplied by 1.4 to account for calibration errors described in \citet{Westfall2019}. See Section~5.8 in \citet{swimdr1} for more information. 

All maps have units of erg s$^{-1}$ cm$^{-2}$ Hz$^{-1}$ arcsec$^{-2}$, except for the mask, which uses 0 to indicate a science-ready pixel and 1 otherwise. The structure of the HDU is given in Table~\ref{table:d4000_dm}.\\

\begin{deluxetable}{cll}[h]
  \tablecaption{HDU0: D4000 Channel Description\label{table:d4000_dm}}
  \renewcommand{\arraystretch}{1.1}
\tablehead{
{Channel} &{Name} & {Description}}
\startdata
{0} & \texttt{Fnu Red} & {Flux density per unit wavelength in the red window}\\
{1} & \texttt{Fnu Blue} & {Flux density per unit wavelength in the blue window}\\
{2} & \texttt{Sigma Red} & {Uncertainty in flux density in the red window}\\
{3} & \texttt{Sigma Blue} & {Uncertainty in flux density in the blue window}\\
{4} & \texttt{Mask}  & D$_n$(4000) mask\\
\enddata
\end{deluxetable}

\noindent\textbf{HDU 1-8: SPECINDX} -- These HDUs contain the information needed to calculate the remaining spectral indices included in SwiM\_v4.1, i.e., the first 43 indices listed in \citet{Westfall2019}. The spectral index D$_n$(4000) is presented in HDU 0 due to its different definition and units. The spectral indices can be calculated using the following equation:
\begin{equation}\label{eqn:app_spidx}
     I_{a} =
    \left\{
	   \begin{array}{ll}
	    	{\Delta\lambda-\frac{F_I}{F_{\rm C0}}} & \mbox{for \AA\ units} \\
		    -2.5\log_{10}[\frac{1}{\Delta\lambda}\frac{F_I}{F_{\rm{ C0}}}] & \mbox{for magnitude units}
	   \end{array}
    \right.
\end{equation}

This relation is described in more detail in Section~5.4 of \cite{swimdr1}. HDUs 1 and 3 contain the flux and the uncertainty maps for the index flux $F_I$ in units of erg s$^{-1}$ cm$^{-2}$ arcsec$^{-2}$, while HDUs 2 and 4 contain the same information for the continuum flux density $F_{\rm C0}$ in units of erg s$^{-1}$ cm$^{-2}\;$\AA$^{-1}$ arcsec$^{-2}$. HDU 5 is the mask for the spectral index maps, where 0 denotes science-ready pixels, and 1 denotes otherwise. 

Each HDU contains a 3D array with the third dimension having a length of 43, corresponding to the 43 included indices. The channel-to-index mapping is provided in the header and in Table~\ref{table:specindx_dm}. We also include the $\Delta \lambda$ for each index in Table~\ref{table:specindx_dm}; this is needed to compute the final indices using Equation~\ref{eqn:app_spidx}.  The index bandpasses are identical to that given in Table 4 of \citet{Westfall2019}.

 HDUs 6--8 contain the flux weighted combined stellar velocity dispersion and instrumental resolution maps, its uncertainty, and mask. The data in HDUs 6 and 7 are in units of km s$^{-1}$, while the masks in HDU 8 have the same definitions as that of HDU 5. These HDUs also have 43 channels corresponding to the 43 indices as given in their header and in  Table~\ref{table:specindx_dm}.\\
 
\begin{deluxetable}{cll}[h]
{\centering
  \caption{HDU 1--8: Spectral Index Channels Description}\label{table:specindx_dm}}
\setlength{\tabcolsep}{30pt}
\renewcommand{\arraystretch}{1.1}
\tablewidth{0pt}
\tablehead{
{Channel} &{Name} & {$\Delta\lambda$\tablenotemark{a} (\AA)}}
\startdata
{0} &  \texttt{CN1} & {35}\\
{1} &  \texttt{CN2} & {35}\\
{2} &  \texttt{Ca4227} & {12.5}\\
{3} &  \texttt{G4300} & {35}\\
{4} &  \texttt{Fe4383} & {51.25}\\
{5} &  \texttt{Ca4455} & {22.5}\\
{6} &  \texttt{Fe4531} & {45}\\
{7} &  \texttt{C24668} & {86.25}\\
{8} &  \texttt{H$\beta$} & {28.75}\\
{9} &  \texttt{Fe5015} & {76.25}\\
{10} & \texttt{Mg1} & {65}\\
{11} & \texttt{Mg2} & {42.5}\\
{12} & \texttt{Mgb} & {32.5}\\
{13} & \texttt{Fe5270} & {40}\\
{14} & \texttt{Fe5335} & {40}\\
{15} & \texttt{Fe5406} & {27.5}\\
{16} & \texttt{Fe5709} & {23.75}\\
{17} & \texttt{Fe5782} & {20}\\
{18} & \texttt{NaD} & {32.5}\\
{19} & \texttt{TiO1} & {57.5}\\
{20} & \texttt{TiO2} & {82.5}\\
{21} & \texttt{H$\delta_\textrm{A}$} & {38.75}\\
{22} & \texttt{H$\gamma_\textrm{A}$} & {43.75}\\
{23} & \texttt{H$\delta_\textrm{F}$} & {21.25}\\
{24} & \texttt{H$\gamma_\textrm{F}$} & {21}\\
{25} & \texttt{CaHK} & {104}\\
{26} & \texttt{CaII1} & {29}\\
{27} & \texttt{CaII2} & {40}\\
{28} & \texttt{CaII3} & {40}\\
{29} & \texttt{Pa17} & {13}\\
{30} & \texttt{Pa14} & {42}\\
{31} & \texttt{Pa12} & {42}\\
{32} & \texttt{MgICvD} & {55}\\
{33} & \texttt{NaICvD} & {28}\\
{34} & \texttt{MgIIR} & {15}\\
{35} & \texttt{FeHCvD} & {30}\\
{36} & \texttt{NaI} & {65.625}\\
{37} & \texttt{bTiO} & {41.5}\\
{38} & \texttt{aTiO} & {155}\\
{39} & \texttt{CaH1} & {44.25}\\
{40} & \texttt{CaH2} & {125}\\
{41} & \texttt{NaISDSS} & {20}\\
{42} & \texttt{TiO2SDSS} & {82.5}\\
\enddata
\tablenotetext{a}{$\Delta\lambda$ is the width of the index band.}
\end{deluxetable}

\clearpage
\noindent\textbf{HDU 9--14: ELINE\_FLUX and ELINE\_EW} -- These HDUs contain the emission line flux and EW maps, and their associated uncertainties. The fluxes come from the Gaussian-fitted measurements from the MPL-11 DAP\null. Each HDU contains a 3D array with the third dimension corresponding to different emission line channels. The channel-to-line mappings are listed in the header and in Table~\ref{table:flux_dm}. 

HDUs 9 and 10 contain the measured flux and uncertainty in units of $10^{-17}$ erg s$^{-1}$ cm$^{-2}$ arcsec$^{-2}$, while HDUs 12 and 13 contain the EW information in units of \AA\null. HDUs 11 and 14 contain the masks for flux and EW, respectively, defined so that 0 denotes science-ready pixels and 1 denotes otherwise.\\

\begin{deluxetable}{cll}[h]
  \caption{HDU 9--14: Emission Line Channels Description\label{table:flux_dm}}
\setlength{\tabcolsep}{30pt}
\renewcommand{\arraystretch}{1.1}
\tablehead{
{Channel} &{Ion} & {$\lambda_{\textrm{rest}}$\tablenotemark{a} (\AA)}}
\startdata
{0} &  \texttt{[\ion{O}{2}]} & {3727.092}\\
{1} &  \texttt{[\ion{O}{2}]} & {3729.875}\\
{2} & \texttt{H12} & {3751.2174}\\
{3} & \texttt{H11} & {3771.7012}\\
{4} &  \texttt{H$\theta$} & {3798.9757}\\
{5} &  \texttt{H$\eta$} & 3836.4720\\
{6} &  \texttt{[\ion{Ne}{3}]} & 3869.86\\
{7} &  \texttt{\ion{He}{1}} & 3889.749\\
{8} &  \texttt{H$\zeta$} & 3890.1506\\
{9} & \texttt{[\ion{Ne}{3}]} & 3968.59\\    
{10} & \texttt{H$\epsilon$} & 3971.1951\\
{11} & \texttt{H$\delta$} & 4102.8922\\
{12} & \texttt{H$\gamma$} & 4341.6837\\
{13} & \texttt{\ion{He}{2}} & 4687.015\\
{14} & \texttt{H$\beta$} & 4862.6830\\
{15} & \texttt{[\ion{O}{3}]} &4960.295\\
{16} & \texttt{[\ion{O}{3}]} & 5008.240\\
{17} & \texttt{[\ion{N}{1}]} & 5199.349\\
{18} & \texttt{[\ion{N}{1}]} & 5201.705\\
{19} & \texttt{\ion{He}{1}} & 5877.252\\
{20} & \texttt{[\ion{O}{1}]} & 6302.046 \\
{21} & \texttt{[\ion{O}{1}]} & 6365.536  \\
{22} &  \texttt{[\ion{N}{2}]}  & 6549.86\\
{23} & \texttt{H$\alpha$} & 6564.608\\
{24} & \texttt{[\ion{N}{2}]}  & 6585.27\\
{25} & \texttt{[\ion{S}{2}]}  &6718.295 \\
{26} & \texttt{[\ion{S}{2}]}  &6732.674\\
{27} & \texttt{\ion{He}{1}} &7067.144 \\
{28} & \texttt{[\ion{Ar}{3}]} & 7137.76\\
{29} & \texttt{[\ion{Ar}{3}]} & 7753.24\\
{30} & \texttt{P$\eta$} & 9017.384\\
{31} & \texttt{[\ion{S}{3}]} & 9071.1\\
{32} & \texttt{P$\zeta$} & 9231.546\\
{33} & \texttt{[\ion{S}{3}]}  & 9533.2\\
{34} &\texttt{P$\epsilon$} & 9548.588\\          
\enddata
\tablenotetext{a}{Vacuum rest wavelengths presented here are from the National Institute of Standards and Technology (NIST) and are used by the MaNGA DAP.}
\end{deluxetable}

\noindent\textbf{HDU 15 --16: SWIFT/SDSS} -- HDU15 contains the sky-subtracted NUV images from \swift\ and optical images from SDSS\null. HDU16 contains their corresponding uncertainty images. All images are in units of nanomaggies. To convert these maps to the AB magnitude ($m$) system, use $m = 22.5 - 2.5\log_{\textrm{10}}(f/\textrm{nanomaggie})$. To convert to $\mu$Jy use 1 $\textrm{nanomaggie}= 3.631$~$\mu$Jy. 

We provide masks for all Swift images in HDU 17 as discussed in Section~\ref{ssec:sw_pipeline}. We strongly recommend users always use the masks from HDU 17 when working with \swift\ images. If there are no bad pixels, then the mask will not change the image. For SDSS, masked pixels have an uncertainty of 0. 

In these HDUs, the data are given in 3D arrays with the third dimension corresponding to the different filters. Their correspondence are given in the header and in Table~\ref{table:phot_dm}.\\

\begin{deluxetable}{clc}[h]
  \tablecaption{HDU 15--16: Photometry Channel Description\label{table:phot_dm}}
\renewcommand{\arraystretch}{1.1}
\tablehead{
{Channel} &{Name} & {Central Wavelength (\AA)}}
\startdata
{0} & \texttt{uvw2} & {1928}\\
{1} & \texttt{uvw1} & {2600}\\
{2} & \texttt{uvm2} & {2246}\\
{3} & \texttt{SDSS u} & {3543}\\
{4} & \texttt{SDSS g}  & {4770}\\
{5} & \texttt{SDSS r} & {6231}\\
{6} & \texttt{SDSS i} & {7625}\\
{7} & \texttt{SDSS z} & {9134}\\
\enddata
\end{deluxetable}\vspace{-10mm}
\noindent\textbf{HDU 17: SWIFT\_UVOT} -- This HDU contains the \swift/UVOT counts, uncertainty, exposure, and mask maps for all three NUV filters. The masks have a value of 0 for science-ready pixels and 1 otherwise. Unlike HDU 15 and 16, these images are {\it not sky-subtracted}. We report the calculated sky counts in the header for each filter under the keywords \texttt{SKY\_W1}, \texttt{SKY\_M2} and \texttt{SKY\_W2}, respectively. The AB magnitude system zero points of the filters and $f_{\lambda}$ conversion factors are also provided in the header as \texttt{ABZP\_}$\ast$ and \texttt{FLAMBDA\_}$\ast$, where the $\ast$ represents the desired filter. 
The structure of this HDU is given in Table~\ref{table:rawswft_dm}.

\begin{deluxetable}{cll}[h]
  \tablecaption{HDU 17: Swift/UVOT Channel Description\label{table:rawswft_dm}}
  \renewcommand{\arraystretch}{1.1}
\tablehead{
{Channel} &{Name} & {Description}}
\startdata
{0} & \texttt{uvw2 Counts} & {Fully reduced, non-sky subtracted uvw2 counts}\\
{1} & \texttt{uvw1 Counts} & {Fully reduced, non-sky subtracted uvw1 counts}\\
{2} & \texttt{uvm2 Counts} & {Fully reduced, non-sky subtracted uvm2 counts}\\
{3} & \texttt{uvw2 Counts Err} & {Uncertainty associated with uvw2 counts}\\
{4} & \texttt{uvw1 Counts Err} & {Uncertainty associated with uvw1 counts}\\
{5} & \texttt{uvm2 Counts Err} & {Uncertainty associated with uvm2 counts}\\
{6} & \texttt{uvw2 Exposure} & {Exposure map for uvw2 image}\\
{7} & \texttt{uvw1 Exposure} & {Exposure map for uvw1 image}\\
{8} & \texttt{uvm2 Exposure} & {Exposure map for uvm2 image}\\
{9} & \texttt{uvw2 Mask} & {Mask for uvw2 image}\\
{10} & \texttt{uvw1 Mask} & {Mask for uvw1 image}\\
{11} & \texttt{uvm2 Mask} & {Mask for uvm2 image}\\
\enddata
\end{deluxetable}

\clearpage
\section{SwiM\_\MakeLowercase{v}4.1 Emission-Line Detection Data Model}\label{app:swimeline}

This appendix provides the data model for the SwiM\_eline\_ratios file, as described in Section \ref{ssec:dr2prod} of this work. This file is new to SwiM\_v4.1, and contains ratios of unmasked-to-total pixels within one elliptical Petrosian radius for each emission line map for each galaxy. The names and contents of each extension are found in Table \ref{table:emline_ratios_dm}.

\begin{deluxetable}{ll}[h]
  \tablecaption{Emission Line Ratios Data Model \label{table:emline_ratios_dm}}
\setlength{\tabcolsep}{30pt}
\renewcommand{\arraystretch}{1.1}
\tablehead{
{Column} & {Description\tablenotemark{a}}}
\startdata
\texttt{MANGAID} & MaNGA ID for the object\\
\texttt{OII-3727} & [\ion{O}{2}] $\lambda3727$\\
\texttt{OII-3729} & [\ion{O}{2}] $\lambda3729$\\
\texttt{H12-3751} & H12 $\lambda3751$\\
\texttt{H11-3771} & H11 $\lambda3771$\\
\texttt{Hthe-3798} &    H$\theta$ $\lambda3798$\\
\texttt{Heta-3826} &    H$\eta$ $\lambda3826$\\
\texttt{NeIII-3869} &    [\ion{Ne}{3}] $\lambda3869$\\
\texttt{HeI-3889} &    \ion{He}{1} $\lambda3889$\\
\texttt{Hzet-3890} &    H$\zeta$ $\lambda3890$\\
\texttt{NeIII-3968} &    [\ion{Ne}{3}] $\lambda3968$\\    
\texttt{Heps-3971} &   H$\epsilon$ $\lambda3971$\\
\texttt{Hdel-4102} &   H$\delta$ $\lambda4102$\\
\texttt{Hgam-4341} &   H$\gamma$ $\lambda4341$\\
\texttt{HeII-4687} &   \ion{He}{2} $\lambda4687$\\
\texttt{Hb-4682} &   H$\beta$ $\lambda4682$\\
\texttt{OIII-4960} &   [\ion{O}{3}] $\lambda4960$\\
\texttt{OIII-5008} &   [\ion{O}{3}] $\lambda5008$\\
\texttt{NI-5199} &   [\ion{N}{1}] $\lambda5199$\\
\texttt{NI-5201} &   [\ion{N}{1}] $\lambda5201$\\
\texttt{HeI-5877} &   \ion{He}{1} $\lambda5877$\\
\texttt{OI-6302} &   [\ion{O}{1}] $\lambda6302$\\
\texttt{OI-6365} &   [\ion{O}{1}] $\lambda6365$\\
\texttt{NII-6549}  &   [\ion{N}{2}] $\lambda6549$\\
\texttt{Ha-6564} &   H$\alpha$ $\lambda6564$\\
\texttt{NII-6585}  &   [\ion{N}{2}] $\lambda6585$\\
\texttt{SII-6718}  &    [\ion{S}{2}] $\lambda6718$\\
\texttt{SII-6732}  &   [\ion{S}{2}] $\lambda6732$\\
\texttt{HeI-7067} &   \ion{He}{1} $\lambda7067$ \\
\texttt{ArIII-7137} &   [\ion{Ar}{3}] $\lambda7137$\\
\texttt{ArIII-7753} &   [\ion{Ar}{3}] $\lambda7753$\\
\texttt{Peta-9017} &   P$\eta$ $\lambda9017$\\
\texttt{SIII-9071} &   [\ion{S}{3}] $\lambda9071$\\
\texttt{Pzet-9231} &   P$\zeta$ $\lambda92331$\\
\texttt{SIII-9533}  &   [\ion{S}{3}] $\lambda9533$\\
\texttt{Peps-9548} &   P$\epsilon$ $\lambda9548$\\          
\enddata
\tablenotetext{a}{Every column except \texttt{MANGAID} presents the ratio of unmasked-to-total pixels within one elliptical Petrosian aperture for the listed emission line.}
\end{deluxetable}

\clearpage

\bibliography{swim}{}
\bibliographystyle{aasjournal}



\end{document}